\documentclass{SCIS2026}
\makeatletter
\renewcommand{\citationline}{%
\@AuthorCitation. \@titlecitation[J]. SCIENCE CHINA Information Sciences, Advance online publication.
DOI: \href{https://doi.org/10.1007/s11432-026-5211-3}{10.1007/s11432-026-5211-3}.%
}
\let\original@maketitle\@maketitle
\renewcommand{\@maketitle}{%
\begin{picture}(0,0)
\put(0,60){\makebox[\textwidth][c]{\small\bfseries Accepted for publication in SCIENCE CHINA Information Sciences}}
\put(0,48){\makebox[\textwidth][c]{\footnotesize\itshape Author's accepted manuscript}}
\end{picture}%
\original@maketitle
}
\makeatother
\begin{document}
\ArticleType{REVIEW}
\Year{2025}
\Month{January}
\Vol{68}
\No{1}
\DOI{}
\ArtNo{}
\ReceiveDate{}
\ReviseDate{}
\AcceptDate{}
\OnlineDate{}
\AuthorMark{}
\AuthorCitation{Shanxing Zeng, Jun Wang, Ying-Chang Liang}

\title{Integrated sensing, communications, and identification for low-altitude economy}{Integrated sensing, communications, and identification for low-altitude economy}

\author[1]{Shanxing ZENG}{}
\author[1]{Jun WANG}{}
\author[2,3]{Ying-Chang LIANG}{{liangyc@ieee.org}}

\address[1]{National Key Laboratory of Wireless Communications,
	University of Electronic Science and Technology of China, \\ Chengdu
	611731, China}
\address[2]{Institute of Fundamental and Frontier Sciences, University of Electronic Science and Technology of China, \\ Chengdu
	611731, China}
\address[3]{Center for Intelligent Networking and Communications,
	University of Electronic Science and Technology of China, \\ Chengdu 611731,
	China}

\abstract{
	With the rapid development of low-altitude economy (LAE), low-altitude activities are becoming increasingly frequent and dynamic, placing growing demands on the underlying information infrastructure to ensure efficient and safe airspace operation. 
	Low-altitude information networks (LAINs), as the key infrastructure supporting these activities, are required not only to ensure reliable communication, but also to support environmental sensing and target identification. 
	Existing research on integrated sensing and communications (ISAC) has established an important foundation for the joint design of communication and sensing functions in LAINs. 
	However, with the increasing need for airspace safety and regulatory compliance, identification is becoming an essential function that should be systematically incorporated into the overall network design. 
	This paper presents a comprehensive review of integrated sensing, communications, and identification (ISCAI) for LAE. 
	We first provide an overview of LAINs by reviewing approaches to low-altitude communication, sensing, and identification, and explain how representative application scenarios motivate their integration. 
	We then review the fundamentals of low-altitude ISAC, including system architectures, channel modeling, unified signal models, and performance metrics, followed by communication-centric and sensing-centric ISAC design as well as resource allocation and aircraft trajectory optimization. 
	Building on these, we introduce the ISCAI paradigm, with particular attention to a backscatter communication (BackCom)-based identification mechanism in which backscatter devices serve as electronic license plates for aircraft. 
	We further present the system and signal models of the resulting integrated sensing and identification (ISAI) and ISCAI paradigms, along with their multifunctional receiver designs and mutual integration gains. 
	Finally, we outline important open problems and future directions, including the information-theoretic framework of ISCAI, artificial intelligence (AI)-empowered ISCAI system design, and reconfigurable intelligent surface-enhanced ISCAI. 
	This paper aims to provide a structured perspective on how LAINs are evolving from the joint design of communication and sensing toward a broader integration of communication, sensing, and identification.
}

\keywords{
	Low-altitude economy (LAE), low-altitude information networks (LAINs), integrated sensing and communications (ISAC), integrated sensing and identification (ISAI), integrated sensing, communications, and identification (ISCAI)
}

\maketitle

\section{Introduction}

Low-altitude economy (LAE) refers to a broad range of economic activities conducted in low-altitude airspace, including unmanned aerial vehicle (UAV) logistics, urban air mobility (UAM), emergency response, and aerial inspection~\cite{re:lae-networks-2025-1,re:lae-networks-2025-2}. 
Driven by advances in UAV technologies, autonomous systems, and airspace management policies, LAE has rapidly evolved from an initial conceptual vision into a highly promising economic domain in recent years~\cite{re:lae-2025-2}. 
Industry forecasts indicate that the global UAV market will reach \$163 billion by 2030~\cite{re:lae-2025-3}, while the global UAM market is expected to reach \$1 trillion by 2040~\cite{re:lae-2021-4}. 
As low-altitude activities continue to grow in scale and diversity, it becomes increasingly important to systematically examine the key elements required to support their efficient and safe operation.

The sustainable operation of LAE depends on three foundational elements, namely low-altitude aircraft, low-altitude airspace, and low-altitude information networks (LAINs).
Low-altitude aircraft, including UAVs, electric vertical takeoff and landing (eVTOL) aircraft, and traditional rotorcraft, serve as the operational entities in LAE and exhibit significant heterogeneity in terms of size, mobility, payload, and mission profiles~\cite{re:lae-aircraft-type-2021}. 
Low-altitude airspace provides the physical environment for low-altitude activities. 
Typically, low-altitude airspace refers to airspace within a vertical distance of 1,000 meters from the ground surface, which can be extended upwards to 3,000 meters according to specific requirements~\cite{re:lae-networks-2025-2,re:lae-networks-2025-7}.
LAINs provide the information infrastructure that connects aircraft with ground facilities and control systems, thereby supporting the efficient and safe operation of low-altitude activities~\cite{re:lae-networks-2025-1,re:lae-networks-2025-8}. Unlike traditional UAV networks that mainly focus on information transmission, LAINs are expected to support a broader set of functions, including communication, sensing, and identification, which are briefly introduced as follows.

\begin{itemize}
	\item
	\textbf{Low-altitude communication:}
	Low-altitude communication provides reliable and low-latency wireless links for both aerial and ground users, thereby supporting flight control, coordination, and mission information transmission~\cite{re:lae-networks-2026-3}. It includes control and non-payload communication (CNPC) such as real-time flight control and navigation commands for eVTOLs, as well as payload communication for service-related data transmission.
	
	\item
	\textbf{Low-altitude sensing:}
	Low-altitude sensing enables real-time awareness of aircraft states and airspace conditions through functions such as localization, velocity estimation, and trajectory tracking~\cite{re:lae-multipoint-2025-1}. By supporting continuous monitoring of targets and surrounding environments, it facilitates key tasks such as obstacle avoidance, trajectory planning, and airspace management.
	
	\item
	\textbf{Low-altitude identification:}
	Low-altitude identification supports the differentiation and recognition of cooperative and non-cooperative targets in shared airspace~\cite{re:lae-multimodal-2026-2}. 
	It includes both target classification and individual identity acquisition. 
	The former is used to separate aircraft from surrounding environmental objects such as birds and to distinguish different types of aircraft, while the latter is the basis for further distinguishing authorized aircraft from unauthorized ones and enabling differentiated control.
\end{itemize}

In low-altitude scenarios, airspace environments are complex and time-varying, with aircraft exhibiting strong three-dimensional (3D) mobility, network topologies changing frequently, and co-channel interference often being severe~\cite{re:lae-networks-2025-1}.
Individual functions alone cannot adequately meet the requirements of complex low-altitude services such as aerial inspection or emergency response, which demand high levels of reliability, safety, and regulatory compliance.
For instance, communication alone can transmit mission-critical information but cannot ensure real-time sensing for obstacle avoidance, while sensing without reliable communication cannot provide timely updates for coordinated operations.
Furthermore, stringent airspace safety regulations require 	continuous identity acquisition and traceability~\cite{re:lae-magazine-2023-isac-7}, tightly coupling identification with communication and sensing.
These conditions collectively reinforce the inherent coupling among communication, sensing, and identification in LAINs, creating strong incentives to treat them as a unified design problem.
Independent designs not only increase resource overhead and mutual interference but also fail to exploit potential integration gains, ultimately constraining overall system performance.

Existing research on integrated sensing and communications (ISAC) has established a solid foundation for the joint design of communication and sensing functions in LAINs. 
A substantial body of work has investigated low-altitude ISAC channel modeling~\cite{re:lae-isac-channel-2025-3}, ISAC waveforms~\cite{re:lae-magazine-2024-isac-9}, communication--sensing performance tradeoffs~\cite{re:lae-magazine-2025-isac-8}, beamforming and resource allocation strategies~\cite{re:lae-isac-2025-2}, and cooperative low-altitude ISAC~\cite{re:lae-multipoint-2025-9}, highlighting the clear advantages of integrated approaches for spectral efficiency and hardware utilization. 
Several magazine and survey articles have also reviewed ISAC for LAINs from different perspectives.
In parallel, dedicated surveys have examined UAV communications and UAV identification as separate research topics.
Table~\ref{tab:low-altitude-isac-sorted-cite} summarizes representative works in these three areas, and compares their coverage of communication, sensing, identification, and other functions~\cite{re:uav-com-2019,re:uav-com-2022,re:lae-multi-ide-2024-1,re:lae-multi-ide-2026-3,re:lae-networks-2025-1,re:lae-networks-2025-2,re:lae-networks-2026-6,re:lae-magazine-2025-isac-5,re:lae-magazine-2025-isac-4,re:lae-magazine-2025-isac-3,re:lae-isac-2026-3,re:lae-clutter-2026-1,re:lae-isac-2026-1,re:lae-isac-2023-4,re:lae-isac-2024-5}.

\begin{table}[!t]
	\footnotesize
	\setlength{\tabcolsep}{4pt}
	\renewcommand{\arraystretch}{1.15}
	\caption{Comparison of representative studies with this work}
	\label{tab:low-altitude-isac-sorted-cite}
	\begin{tabularx}{\textwidth}{p{0.045\textwidth}p{0.032\textwidth}ccc p{0.17\textwidth} >{\raggedright\arraybackslash}X}
		\toprule
		Ref. & Year & Comm. & Sens. & Iden. & Other functions & Main contributions \\
				\midrule
		\multicolumn{7}{@{}l}{\textit{UAV communication surveys}} \\
		\cite{re:uav-com-2019} & 2019 & \checkmark & -- & -- & -- & Provided an in-depth review of UAV-assisted wireless communications and cellular-connected UAV communications.\\
		\cite{re:uav-com-2022} & 2022 & \checkmark & -- & -- & -- &  Reviewed the evolution of UAV cellular communications from 5G to 6G, and discussed corresponding enabling technologies.\\
		\midrule
		\multicolumn{7}{@{}l}{\textit{UAV identification surveys}} \\
		\cite{re:lae-multi-ide-2024-1} & 2024 & -- & \checkmark & \checkmark & -- &Summarized four types of UAV detection and classification technologies based on machine learning algorithms. \\
		\cite{re:lae-multi-ide-2026-3} & 2026 & -- & \checkmark & \checkmark & -- &Reviewed the detection, classification, and tracking of autonomous aerial vehicles using radar and communication systems. \\
		\midrule
		\multicolumn{7}{@{}l}{\textit{ISAC studies for LAINs}} \\
		\cite{re:lae-networks-2025-1} & 2025 & \checkmark & \checkmark & $\triangle$ & Computing, positioning, navigation, control & Presented the core architectures and three categories of enabling technologies for LAINs including the integration of communication, sensing, and computing.\\
		\cite{re:lae-networks-2025-2} & 2026 & \checkmark & \checkmark & $\triangle$ & Computing, control, wireless power transfer & Reviewed the architectures, performance metrics, and multi-layer defense framework of LAINs, and investigated multifunctional system designs such as ISAC. \\
		\cite{re:lae-networks-2026-6} & 2025 & \checkmark & \checkmark & -- & Computing, control & Presented the architectures, applications, and challenges of LAINs. \\
		\cite{re:lae-magazine-2025-isac-5} & 2025 & \checkmark & \checkmark & $\triangle$ & -- & Investigated four key issues of ISAC for LAINs: accurate target detection, interference management, communication--sensing coverage mismatch, and air--ground coordination alongside handover complexity. \\
		\cite{re:lae-magazine-2025-isac-4} & 2025 & \checkmark & \checkmark & -- & -- & Explored cellular-connected and UAV-assisted ISAC paradigms for LAINs, focusing on spectrum sharing, interference cancellation, and cooperative active sensing. \\
		\cite{re:lae-magazine-2025-isac-3} & 2026 & \checkmark & \checkmark & -- & Computing, control, wireless power transfer & Extended ISAC with capabilities in control, computation, wireless power transfer, and large language model-based intelligence for LAINs. \\
		\cite{re:lae-isac-2026-3} & 2026 & \checkmark & \checkmark & -- & -- & Explored full-duplex ISAC for LAINs. \\
		\cite{re:lae-clutter-2026-1} & 2026 & \checkmark & \checkmark & -- & -- & Focused on robust ISAC system design under strong clutter interference for LAINs. \\
		\cite{re:lae-isac-2026-1} & 2026 & \checkmark & \checkmark & -- & -- & Introduced game-theoretic learning approaches to enhance ISAC performance for LAINs. \\
		\cite{re:lae-isac-2023-4} & 2023 & \checkmark & \checkmark & -- & -- & Introduced air--ground ISAC for LAINs, and analyzed the characteristics and advantages of air--ground ISAC. \\
		\cite{re:lae-isac-2024-5} & 2024 & \checkmark & \checkmark & -- & -- & Explored UAV-assisted ISAC for LAINs, and designed various solutions for optimizing ISAC performance. \\
		\midrule
		\textbf{Ours} & \textbf{2026} & \checkmark & \checkmark & \checkmark & -- & \textbf{Incorporates identification into the joint design of communication and sensing, introduces a BackCom-based identification mechanism where backscatter devices (BDs) serve as electronic license plates for aircraft, and presents the system and signal models, multifunctional receiver design, and mutual integration gains of ISAI and ISCAI.} \\
		\bottomrule
	\end{tabularx}
	\vspace{2pt}
	\par\noindent\footnotesize
	Comm.: communication; Sens.: sensing; Iden.: identification.
	\checkmark: substantively reviewed or explicitly incorporated into system design;
	$\triangle$: briefly discussed or included only as a supporting element;
	--: not addressed.
\end{table}

At the standardization level, the 3rd Generation Partnership Project (3GPP) Releases 17 through 19 have progressively enhanced support for UAV and UAM connectivity, addressing coverage improvement, interference management, and identity management for unmanned aerial systems.
In particular, 3GPP TS 23.256 defines frameworks for UAV connectivity, identity acquisition, authentication, and tracking within cellular networks~\cite{re:3gpp_ts23256}, while 3GPP TR 22.837 studies sensing use cases and potential functional requirements for ISAC~\cite{re:3gpp_tr22837}, covering applications such as object detection and tracking, environment monitoring, and motion monitoring. Building on this, 3GPP TS 22.137 formally specifies the service requirements and performance metrics for ISAC-enabled systems, including key indicators such as positioning accuracy and velocity accuracy~\cite{re:3gpp_ts22137}.
IEEE standards have provided guidance on UAV cellular communication terminal specifications and low-altitude airspace structures~\cite{re:ieee1939_1_2021,re:ieee1937_8_2024}.

Despite these advances, a persistent gap remains in both research and standardization: identification is still designed and operated as a standalone module, and has not yet been incorporated into a joint design with communication and sensing.
Table~\ref{tab:low-altitude-isac-sorted-cite} makes the resulting research gap explicit. 
Surveys on UAV communications concentrate on the communication function, and surveys on UAV identification concentrate on identification itself. 
Among existing ISAC studies for LAINs, several works have extended low-altitude ISAC toward computing, control, and wireless power transfer, whereas only a few mention identification as a requirement and none designs identification jointly with communication and sensing.

\begin{figure*}[!t]
	\centering
	\includegraphics[width=0.85\textwidth]{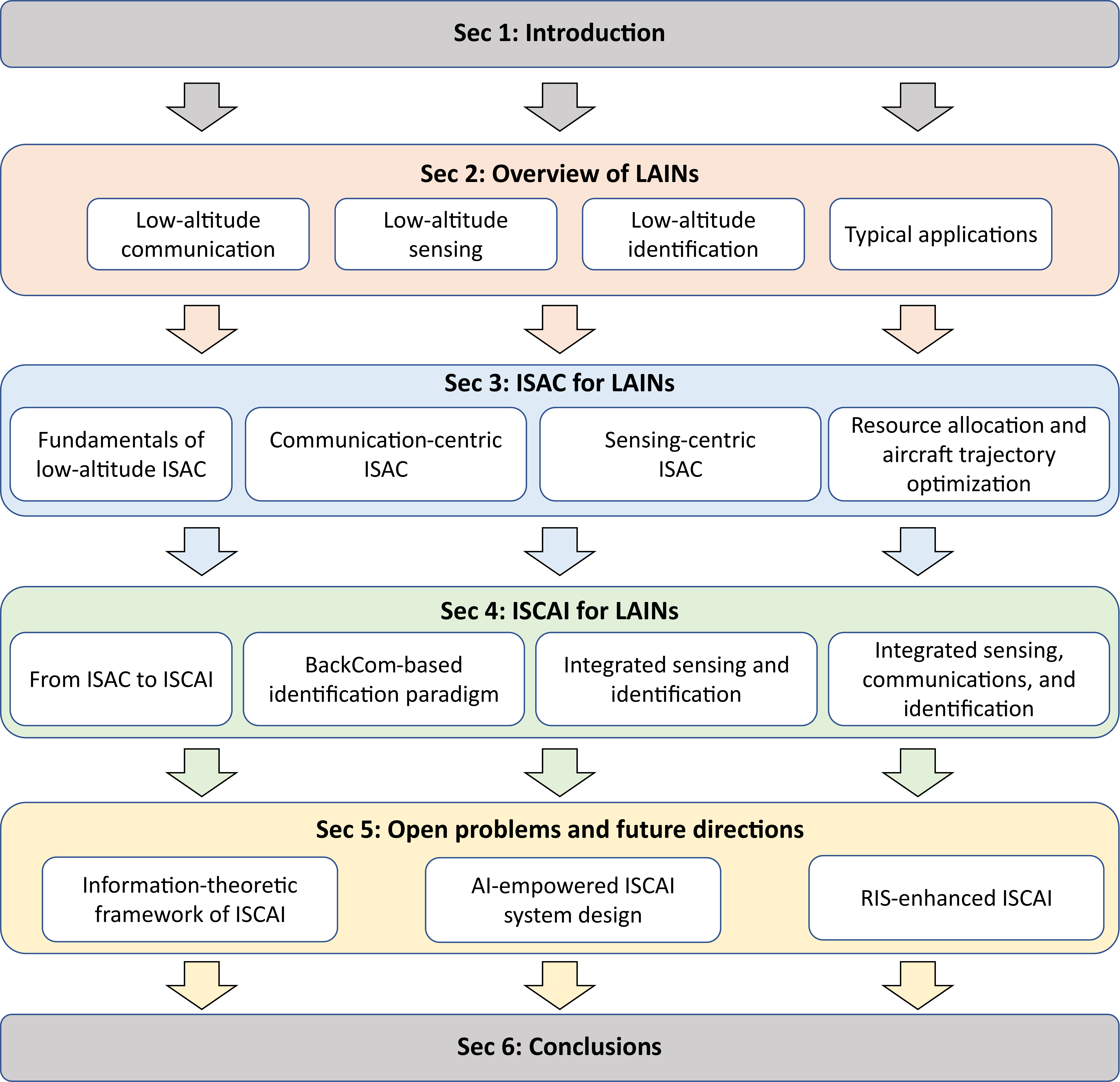}	
	\caption{
	Organization of this paper.
	Section~2 mainly reviews three fundamental functions of LAINs, while Section~3 discusses existing ISAC studies for LAINs.
	Then, Section 4.1 discusses the integration of existing identification paradigms with ISAC and clarifies their limitation in forming ISCAI.
	To cope with this limitation, Section 4.2 introduces BackCom-based identification as a complement to existing identification paradigms and regards it as the key technical bridge from existing ISAC toward ISAI and ISCAI.
	Section 4.3 integrates BackCom-based identification into low-altitude sensing to form ISAI, and Section 4.4 further integrates BackCom-based identification into ISAC to form ISCAI.
	Section 5 outlines open problems and future directions of ISCAI, while Section 6 concludes the paper.
}
	\label{fig:organization}
\end{figure*}

To address this gap, this paper presents a comprehensive review of integrated sensing, communications, and identification (ISCAI) for LAE. 
In this paper, ISCAI refers to a system framework that incorporates identification into the joint design of communication and sensing, enabling data transmission, physical-state sensing, and target identification to be performed within an integrated system.
We begin by reviewing the main standalone approaches to low-altitude communication, sensing, and identification, and explain how representative application scenarios, including smart logistics, UAM, emergency rescue, and low-altitude agriculture, motivate their integration into a unified design framework. 
We then review the fundamentals of low-altitude ISAC, including system architectures, channel modeling, unified signal models, and performance metrics, followed by communication-centric and sensing-centric ISAC design, as well as resource allocation and aircraft trajectory optimization, serving as the technical foundation upon which ISCAI is built. 
Next, we introduce the ISCAI paradigm, with particular attention to a backscatter communication (BackCom)-based identification mechanism where backscatter devices (BDs) serve as electronic license plates for aircraft. We further present the system and signal models of the resulting integrated sensing and identification (ISAI) and ISCAI paradigms, along with their multifunctional receiver design and mutual integration gains. Finally, we discuss three open problems and future research directions, namely the information-theoretic framework of ISCAI, artificial intelligence (AI)-empowered ISCAI system design, and reconfigurable intelligent surface (RIS)-enhanced ISCAI.
For clarity, the main abbreviations used throughout this paper are summarized in Table~\ref{tab:abbreviations}.

As shown in Figure~\ref{fig:organization}, the remainder of this paper is organized as follows.
Section 2 provides an overview of LAINs.
Section 3 reviews low-altitude ISAC.
Section 4 presents the ISAI and ISCAI paradigms.
Section 5 outlines open problems and future directions.
Section 6 concludes the paper.

\begin{table}[!t]
	\centering
	\footnotesize
	\setlength{\tabcolsep}{4pt}
	\renewcommand{\arraystretch}{1.15}
	\caption{List of main abbreviations}
	\label{tab:abbreviations}
	\begin{tabularx}{\textwidth}{@{}l X l X@{}}
		\toprule
		Abbreviation & Description & Abbreviation & Description \\
		\midrule
		2D      & Two-dimensional                                        & 3D      & Three-dimensional                                      \\
		3GPP    & 3rd Generation Partnership Project                     & ADS-B   & Automatic dependent surveillance--broadcast            \\
		AFDM    & Affine frequency division multiplexing                 & AI      & Artificial intelligence                                \\
		AoA     & Angle of arrival                                       & AoD     & Angle of departure                                     \\
		AWGN    & Additive white Gaussian noise                          & BackCom & Backscatter communication                              \\
		BD      & Backscatter device                                     & BER     & Bit error rate                                         \\
		CNPC    & Control and non-payload communication                  & CP      & Cyclic prefix                                          \\
		CRLB    & Cram\'{e}r--Rao lower bound                            & CS      & Compressive sensing                                    \\
		DAFT    & Discrete affine Fourier transform                      & eVTOL   & Electric vertical takeoff and landing                  \\
		FFT     & Fast Fourier transform                                 & FIM     & Fisher information matrix                              \\
		FMCW    & Frequency-modulated continuous wave                    & FSK     & Frequency shift keying                                 \\
		GBSM    & Geometry-based stochastic model                        & ICI     & Inter-carrier interference                             \\
		IDE     & Iterative detection and estimation                     & IoT     & Internet of Things                                     \\
		ISAC    & Integrated sensing and communications                  & ISAI    & Integrated sensing and identification                  \\
		ISCAI   & Integrated sensing, communications, and identification & LAE     & Low-altitude economy                                   \\
		LAINs   & Low-altitude information networks                      & LEO     & Low Earth orbit                                        \\
		LoS     & Line-of-sight                                          & LS      & Least squares                                          \\
		MI      & Mutual information                                     & ML      & Maximum likelihood                                     \\
		MMSE    & Minimum mean square error                              & MSE     & Mean square error                                      \\
		MTI     & Moving target indication                               & MUSIC   & Multiple signal classification                         \\
		OFDM    & Orthogonal frequency division multiplexing             & OMP     & Orthogonal matching pursuit                            \\
		OTFS    & Orthogonal time frequency space                        & PAPR    & Peak-to-average power ratio                            \\
		PMU     & Power management unit                                  & PSK     & Phase shift keying                                     \\
		RCS     & Radar cross section                                    & Remote ID & Remote identification                                  \\
		RF      & Radio frequency                                        & RFF     & Radio frequency fingerprinting                         \\
		RFID    & Radio frequency identification                         & RIS     & Reconfigurable intelligent surface                     \\
		SBL     & Sparse Bayesian learning                               & SINR    & Signal-to-interference-plus-noise ratio                \\
		UAM     & Urban air mobility                                     & UAV     & Unmanned aerial vehicle                                \\
		UPA     & Uniform planar array                                   &         &                                                        \\
		\bottomrule
	\end{tabularx}
\end{table}

\section{Overview of LAINs}
\label{sec:LAINs-infra}
This section provides an overview of LAINs by reviewing the standalone approaches to low-altitude communication, sensing, and identification, together with representative application scenarios.

\subsection{Low-altitude communication}

Low-altitude communication provides reliable and low-latency wireless links for both aerial and ground users, thereby supporting flight control, coordination, and mission data transmission for low-altitude aircraft. 
Unlike traditional terrestrial communication networks designed primarily for ground users, low-altitude communication operates in a fundamentally different environment characterized by 3D airspace extension, highly dynamic aerial mobility, and the dense coexistence of heterogeneous aerial and ground users. 
To support these demands, low-altitude communication can adopt a layered architecture in which terrestrial cellular networks provide the primary communication infrastructure, while satellite networks and aerial ad hoc networks serve as complementary layers that extend connectivity in the geographic and topological dimensions, respectively~\cite{re:uav-com-2019,re:uav-com-2022}.

\subsubsection{Terrestrial cellular network-based low-altitude communication}

Terrestrial cellular networks serve as the primary communication infrastructure for LAINs, owing to their wide deployment, mature protocols, and proven performance in supporting mobile users. 
However, directly applying existing ground-oriented cellular systems to low-altitude scenarios introduces significant performance degradation due to several fundamental mismatches between network design assumptions and low-altitude operational realities.

\textbf{Coverage limitations in 3D airspace:} 
The antennas of existing terrestrial base stations are tilted downward, either mechanically or electrically, so as to maximize coverage on the ground plane. 
As a result, their coverage in the vertical dimension is limited.
Aircraft at altitudes well above the base-station antennas are typically served by antenna sidelobes~\cite{re:uav-com-2019}.
Such aircraft receive weak signals and suffer from unstable connectivity.

\textbf{Severe interference under high line-of-sight (LoS) probability:} 
A distinctive feature of low-altitude channels is their significantly higher LoS probability than that of ground-level channels, mainly due to the reduced blockage caused by buildings and terrain at higher altitudes~\cite{re:uav-com-2019}. 
While high LoS probability improves the quality of the serving link, it simultaneously reduces spatial isolation between co-channel users, exposing aerial users to strong interference from numerous non-serving base stations across wide geographic areas~\cite{huang2026new,cai2019interference}.

\textbf{High channel non-stationarity introduced by aircraft mobility:} 
The 3D mobility of low-altitude aircraft leads to rapid and continuous variations in channel conditions, which significantly complicate channel estimation, tracking, and prediction. 
As aircraft move through the airspace, the angles of arrival (AoAs) and angles of departure (AoDs), path delays, and Doppler frequency shifts of multipath components change at rates far exceeding those encountered in typical ground mobile scenarios, resulting in pronounced channel non-stationarity~\cite{xu20223}. 
For high-speed aircraft, large Doppler shifts cause severe inter-carrier interference (ICI) in classical orthogonal frequency division multiplexing (OFDM)-based communication systems.

To extend vertical coverage, antenna designs with adjustable beam elevation have been investigated, including rotatable antennas that support mechanical or electronic tilt adjustment~\cite{li2025rotatable,zheng2025rotatable} and full-dimension multiple-input multiple-output (MIMO) systems equipped with two-dimensional (2D) planar arrays that enable independent beamforming in both azimuth and elevation dimensions~\cite{re:uav-com-2019}. 
These approaches allow base stations to steer main lobes toward low-altitude airspace while mitigating interference toward neighboring cells.

To manage inter-cell interference, terrestrial cellular networks can exploit knowledge of aircraft flight trajectories to enable proactive interference suppression through multi-domain cooperative resource management. 
The core idea is to coordinate wireless resources across the spatial, power, time, and frequency domains jointly, driven by global channel state information obtained through inter-base-station information exchange, so that the strong and unstructured interference can be shaped into manageable or even constructive signal combinations at the receiver. 
Current research covers joint 3D trajectory and power design for interference management in UAV-assisted networks~\cite{re:lae-inter-2022-2}, coordinated beamforming of terrestrial base stations to protect aerial users~\cite{re:lae-inter-2026-4}, and joint trajectory optimization and adaptive interference cancellation for cellular-connected UAVs~\cite{re:lae-inter-2022-6}.

To cope with high channel non-stationarity, accurate channel modeling forms the necessary foundation for system design. 
Dedicated non-stationary channel models have been developed to characterize the space--time--frequency variations of low-altitude UAV channels, accounting for aircraft altitude, velocity, and moving direction alongside time-varying multipath components~\cite{xu20223}. 
Beyond channel modeling, RIS has been explored as a means of stabilizing channel conditions in highly dynamic environments by constructing virtual LoS links for aircraft, thereby reducing the severity of channel fluctuations experienced by aerial users~\cite{zhang2026ris}.

\subsubsection{Satellite network-assisted low-altitude communication}
Despite the enhancements introduced at the terrestrial network level, ground-based cellular infrastructure alone cannot provide seamless communication coverage for all low-altitude activities.  
Remote mountainous regions, open seas, and rural areas may lack sufficient base station deployment, leaving large geographic areas without reliable connectivity. 
Satellite networks offer a natural complement in these scenarios, leveraging their wide-area coverage and largely unobstructed propagation paths to extend communication coverage into regions and altitudes beyond the reach of ground-based systems~\cite{re:lae-networks-2025-7,gao2026agentic}. 
Among different orbital tiers, low Earth orbit (LEO) satellites are particularly attractive for low-altitude communication support due to their relatively low propagation delay, higher received signal strength, and capability of providing wide-area coverage through large-scale constellations.

In low-altitude communication, satellites provide wide-area or backup connectivity, while terrestrial base stations remain responsible for localized high-capacity access~\cite{cui2022space}. 
However, realizing satellite-assisted low-altitude communication also introduces several technical challenges. 
The high relative mobility between LEO satellites and low-altitude aircraft leads to rapidly varying link geometry, large Doppler shifts, and frequent beam misalignment, which complicate synchronization and link maintenance. 
In addition, when satellite, aerial, and terrestrial networks share the same frequency bands, the resulting cross-network interference requires careful coordination~\cite{zhou2024hierarchical}.

To address these challenges, existing studies have investigated satellite-assisted low-altitude communication from multiple perspectives. 
Dynamic service migration and request routing in time-varying satellite-aided UAV networks have been studied to reduce delay and improve throughput under limited local observations~\cite{jia2025satellite}. 
Secure transmission in integrated satellite--UAV low-altitude networks has also attracted increasing attention, where physical-layer security mechanisms have been introduced to protect low-altitude wireless links in open propagation environments~\cite{li2025securing}. 
For large-scale low-altitude Internet of Things (IoT) services, grant-free random access schemes tailored to satellite-assisted networks have been developed to improve access efficiency and the reliability of activity detection and channel estimation~\cite{pei2025expanded}. 
Moreover, communication-aware UAV trajectory design and data collection strategies have been studied in LEO-satellite-assisted systems to reduce energy consumption while satisfying heterogeneous transmission requirements~\cite{jia2020leo}. 
These works indicate that research on satellite-assisted low-altitude communication has gone beyond basic coverage extension and is gradually moving toward more integrated designs involving routing, security, massive access, and mobility optimization.

\subsubsection{Aerial ad hoc network-based low-altitude communication}
Terrestrial cellular and satellite networks mainly provide links between aircraft and infrastructure, so information exchange among aircraft is typically relayed through an ``aircraft--infrastructure--aircraft'' path.
For cooperative scenarios such as formation flight, cooperative obstacle avoidance, and swarm task allocation, this indirect path not only introduces additional round-trip latency, but also causes the whole cooperation capability to fail once the infrastructure suffers coverage holes or link outages.
Direct air-to-air communication establishes direct links among aircraft, allowing cooperative information to be exchanged locally in the air, and enables aircraft at the edge of coverage to access the network through multi-hop relaying, thereby complementing the coverage provided by the terrestrial and satellite layers in a topology-extension manner~\cite{re:lae-a2a-2019-1,re:lae-a2a-2025-2}.

However, such infrastructure-free networking also introduces new challenges.
First, both ends of the link are highly mobile, so the network topology changes continuously, communication links have short lifetimes, routes need to be frequently reconstructed, and the control overhead grows rapidly with the number of nodes.
Second, in fully distributed configurations without a central scheduler, coordinating channel access is challenging, and hidden-terminal effects and access collisions can become pronounced in dense formations.
In addition, the energy consumed by relaying competes directly with flight endurance, and the onboard radio frequency (RF) capability is also constrained by the platform payload, which further limits the communication range and data rate that air-to-air links can support.

To address these challenges, existing studies have investigated air-to-air communication from multiple perspectives.
The modeling and measurement of air-to-air channels provide the basis for link budget analysis and transmission parameter configuration of air-to-air links~\cite{re:lae-a2a-2019-1}.
At the networking level, position- and trajectory-aided predictive routing establishes backup paths in advance by estimating link lifetimes~\cite{re:lae-a2a-2020-3}.
In terms of access and resource management, cooperative spectrum sensing can be used to detect spectrum occupancy~\cite{re:lae-spe-2025-5}, sidelink-based autonomous resource selection allows aircraft to access the channel without centralized scheduling~\cite{re:3gpp_ts24577}, and multi-agent reinforcement learning has been used to optimize relay selection and power allocation in UAV swarms based on local and neighbor-shared observations~\cite{re:lae-a2a-2023-4}.
Given that the mobility of aircraft is itself controllable, formation-aware distributed topology control adjusts UAV positions to improve access-link quality, inter-UAV gateway connectivity, and load distribution while maintaining a connected topology~\cite{re:lae-a2a-2025-5}.
These works show that research on aerial ad hoc network-based low-altitude communication has gradually expanded from the early networking problem centered on connectivity guarantee to multiple aspects including channel characterization, distributed resource management, and coupling with flight control.

\subsection{Low-altitude sensing}

Low-altitude sensing enables real-time awareness of aerial target states and airspace conditions, with core tasks including target detection, localization, velocity estimation, and trajectory tracking. 
In low-altitude airspace, sensing targets are highly heterogeneous, ranging from large eVTOL aircraft and manned rotorcraft to small consumer-grade UAVs, which exhibit distinct physical sizes, flight speeds, and motion patterns. 
Among existing sensing technologies, radar remains the dominant solution for low-altitude airspace surveillance due to its all-weather capability, long sensing range, and ability to estimate multiple target parameters without requiring target cooperation~\cite{re:hu2019improved,re:lae-multi-ide-2026-3}.

\textbf{Monostatic radar:} 
In monostatic configurations, the transmitter and receiver are co-located, forming a self-contained sensing unit that estimates target range, radial velocity, and angle from reflected echoes. 
Frequency-modulated continuous wave (FMCW) radar is widely adopted in low-altitude sensing due to its high range resolution, low transmit power, and compact hardware footprint, making it suitable for deployment on both ground stations and aerial platforms~\cite{re:lae-radar-2018-2}. 
For small UAV sensing, micro-Doppler analysis provides a means of extracting fine-grained rotor motion parameters from the periodic frequency modulation induced by rotating blades, enriching the physical state information available beyond simple range and velocity estimates~\cite{re:lae-isac-spc-2025-1}. 
However, monostatic radar is inherently constrained by single-perspective observation, which makes it more vulnerable to coverage blind spots caused by building or terrain blockage.

\textbf{Multistatic and networked radar:} 
Multistatic radar systems employ spatially separated transmitters and receivers to observe the same target from multiple geometric perspectives, thereby providing spatial diversity that cannot be achieved by single-station configurations. By fusing echoes received at multiple nodes, multistatic systems significantly improve parameter estimation accuracy, resolve ambiguities arising from multipath propagation, and maintain continuous tracking of targets that enter the coverage blind spots of individual stations~\cite{chernyak2018fundamentals}. Networked radar further extends this concept to large-scale cooperative architectures with geographically distributed sensing nodes, enabling wide-area monitoring and more robust trajectory reconstruction in complex urban environments~\cite{zhou2023networked}. Key research challenges include time--frequency synchronization among distributed nodes~\cite{hochberg2026time}, multi-target data association and track fusion across stations~\cite{sruti2024nonparametric}, and joint resource allocation for coordinated illumination and reception~\cite{sun2020resource}.

\textbf{Passive radar:} 
Passive radar exploits ambient RF signals from non-cooperative illuminators of opportunity, including frequency-modulation broadcast signals, digital television signals, and cellular signals, as sensing carriers, eliminating the need for dedicated transmitters and enabling covert operation~\cite{griffiths2022introduction}. 
In low-altitude sensing, passive radar based on cellular signals is particularly attractive because dense cellular deployments in urban areas can provide broad spatial coverage and diverse illumination directions for target localization and tracking. 
By correlating the direct-path reference signal with target-reflected echoes, passive radar systems can estimate bistatic range and Doppler information. 
Nevertheless, their performance depends on the availability, waveform characteristics, and geometry of the illuminators, and is constrained by weak target echoes, strong direct-path interference, and limited control over the transmitted signals~\cite{colone2022passive}.

In addition to radar, other sensing modalities can also complement low-altitude airspace perception in specific scenarios. 
Optical and infrared sensors provide rich spatial and thermal information for short-range observation~\cite{re:lae-multi-ide-2025-7}, light detection and ranging (LiDAR) supports high-precision obstacle perception and local environmental mapping~\cite{khan2021comparative}, and acoustic sensing can assist passive monitoring by exploiting target sound signatures~\cite{chen2024low}. 
However, compared with radar, these modalities are generally more sensitive to illumination, weather, environmental noise, or sensing range limitations.

\subsection{Low-altitude identification}
Low-altitude identification supports the recognition and differentiation of aerial targets in shared airspace, and is essential for airspace supervision, regulatory compliance, and security management. 
Unlike sensing, which focuses on estimating physical parameters such as position, velocity, and trajectory, identification aims to determine either what the target is or who the target is.
Accordingly, low-altitude identification can be broadly divided into two categories, namely target classification and individual identity acquisition. 
The former distinguishes different types of aerial objects, while the latter aims to acquire the individual identifier associated with a target platform.

\subsubsection{Target classification}
Target classification aims to distinguish different aerial objects, including various UAV types, manned aircraft, birds, and other airborne entities, based on observable physical or signal characteristics~\cite{re:lae-multi-ide-2026-3}. Existing approaches mainly rely on features extracted from RF signals, radar echoes, images, videos, or acoustic signatures~\cite{re:lae-multi-ide-2024-2}.

\textbf{RF signal-based classification} 
captures and analyzes RF signals actively transmitted by targets, and exploits features such as time--frequency characteristics, waveform structures, and protocol signatures to infer the target type~\cite{re:lae-multi-ide-2022-10}. This approach is effective for targets that are actively transmitting, but its applicability is inherently limited when the target remains silent or adopts unknown transmission formats.

\textbf{Vision-based classification} 
relies on images or video streams collected by optical sensors, and typically leverages deep learning models to identify targets from their shapes, textures, and motion patterns~\cite{re:lae-multi-ide-2022-11}. It can provide rich spatial information and intuitive interpretability, but its performance is highly sensitive to illumination conditions, weather, and background clutter.

\textbf{Acoustic signal-based classification} 
exploits the sound signatures generated by UAV motors and propellers, and is particularly useful for short-range detection and recognition~\cite{re:lae-multi-ide-2019-13}. 
However, it is vulnerable to ambient noise and wind interference, and is further constrained by limited sensing range, which limits its robustness in practical outdoor environments.

\textbf{Radar-based classification} 
uses echo features such as radar cross section (RCS) and micro-Doppler signatures to distinguish target types~\cite{11192227}, and is especially attractive for non-cooperative targets because it does not rely on target transmission or favorable illumination conditions~\cite{re:lae-multi-ide-2024-12}. Nevertheless, weak echoes from small UAVs and strong environmental clutter can make reliable feature extraction challenging.

Since each single-modality method has inherent limitations, multi-modal classification has become an important direction for improving target classification accuracy in low-altitude environments~\cite{re:lae-multi-ide-2025-6,re:lae-multi-ide-2025-7}. By fusing complementary information from heterogeneous sensing sources, multi-modal approaches can achieve more robust and accurate classification across diverse operating conditions. For example, a system may combine radar sensing, which is robust to weather and illumination variations, with visual sensing, which provides rich appearance information, so that the two modalities can compensate for each other in target recognition. However, such gains also come with additional challenges, including cross-modal synchronization, heterogeneous data alignment, and increased processing complexity.

\subsubsection{Individual identity acquisition}

Individual identity acquisition goes beyond target type recognition and aims to acquire the specific registered identity of each aircraft~\cite{re:3gpp_ts23256}. 
Existing approaches can be broadly grouped into three paradigms, namely broadcast-based identity reporting, network-based identity reporting, and radio frequency fingerprinting (RFF).

\textbf{Broadcast-based identity reporting} includes technologies such as automatic dependent surveillance--broadcast (ADS-B) and broadcast remote identification (Remote ID)~\cite{re:lae-multi-ide-2020-9,re:lae-multi-ide-2024-8}. In these schemes, aircraft periodically broadcast their identity, position, and status information, which can be passively received by nearby infrastructure or other airspace users. Their main advantage lies in simple implementation and low dependence on network infrastructure. However, such schemes rely entirely on target cooperation and truthful self-reporting. As a result, they are vulnerable to spoofing, forgery, or intentional shutdown, and thus cannot always provide trustworthy identity acquisition in contested or unregulated low-altitude environments.

\textbf{Network-based identity reporting} requires aircraft to connect to terrestrial communication networks and upload identifiers, flight plans, and operational data to network-side management platforms~\cite{re:3gpp_ts23256}. Compared with broadcast-based approaches, this paradigm enables stronger authentication, centralized supervision, and tighter integration with airspace management systems. Nevertheless, its effectiveness depends on the availability and reliability of communication connectivity, as well as the active reporting of aircraft. For example, in remote areas, coverage-deficient regions, or disrupted environments, network-based identity uploading may become unavailable or unstable, which limits its applicability as a universal identification solution.

\textbf{Radio frequency fingerprinting} identifies aircraft by exploiting device-specific hardware imperfections embedded in transmitted RF signals~\cite{re:lae-rff-2020-1}. These features arise from manufacturing tolerances in RF front-end components and are therefore difficult to replicate exactly, allowing them to serve as device-level identifiers~\cite{re:lae-rff-2024-2}. 
This makes RFF a useful non-cooperative complement to self-reporting-based identification schemes, especially when the target is suspicious or unwilling to disclose its identity. 
However, RFF-based identification necessitates active signal transmission from targets, and it is rendered ineffective when targets maintain radio silence.
Moreover, its performance depends heavily on the quality of the fingerprint database and can be affected by channel variations, device aging, and signal quality, which makes robust deployment in dynamic low-altitude environments still challenging.

Overall, each paradigm addresses only part of the identification problem. 
Broadcast-based and network-based approaches are more suitable for cooperative and connected aircraft, whereas RFF provides a complementary solution for non-cooperative or suspicious targets. 
These limitations of existing schemes also motivate the development of more controllable identification mechanisms, which will be further discussed in Section 4.

\subsection{Typical applications for LAINs}

As shown in Table~\ref{tab:low-altitude-requirements}, LAINs encompass a wide range of application scenarios with distinct operational characteristics and performance requirements~\cite{re:lae-application-2025-1,re:lae-whitepaper-2025,re:3gpp_ts22125}. Representative scenarios such as smart logistics, UAM, emergency rescue, and low-altitude agriculture impose different yet closely coupled demands on communication, sensing, and identification, thereby motivating their integrated support in LAINs.

\begin{table*}[!b]
	\footnotesize
	\caption{Requirement comparison of representative low-altitude application scenarios~\cite{re:lae-whitepaper-2025,re:3gpp_ts22125}}
	\label{tab:low-altitude-requirements}
	\centering
	\setlength{\tabcolsep}{4pt}
	\renewcommand{\arraystretch}{1.12}
	\begin{tabular}{p{0.17\textwidth}p{0.23\textwidth}p{0.26\textwidth}p{0.26\textwidth}}
		\toprule
		Application scenario & Communication requirements & Sensing requirements & Identification requirements \\
		\midrule
		Low-altitude smart logistics
		&
		$\bullet$ Data rate of tens to hundreds of kbps for control links. \newline
		$\bullet$ Payload data rate of several Mbps. \newline
		$\bullet$ End-to-end latency $<100$ ms.
		&
		$\bullet$ Obstacle ranging accuracy from centimeters to meters. \newline
		$\bullet$ Continuous tracking of active delivery UAVs for traffic management.
		&
		$\bullet$ Real-time identity acquisition of registered UAVs. \newline
		$\bullet$ Detection of unauthorized intruders.
		\\
		\midrule
		Urban air mobility
		&
		$\bullet$ End-to-end latency $<10$ ms. \newline
		$\bullet$ Availability $>99.999\%$. \newline
		$\bullet$ Highly reliable CNPC links for safety-critical operations.
		&
		$\bullet$ Positioning accuracy at the decimeter level. \newline
		$\bullet$ Real-time tracking of airspace users and unexpected obstacles.
		&
		$\bullet$ Unique and continuous identity acquisition of each eVTOL. \newline
		$\bullet$ Detection of identity spoofing and unauthorized aircraft.
		\\
		\midrule
		Low-altitude emergency rescue
		&
		$\bullet$ End-to-end latency $<10$ ms. \newline
		$\bullet$ Reliability $>99.999\%$. \newline
		$\bullet$ Robust links in impaired environments.
		&
		$\bullet$ Survivor localization accuracy of 1--3 m. \newline
		$\bullet$ Rapid area scanning using multi-modal sensing.
		&
		$\bullet$ Identification of rescue UAVs under degraded network conditions. \newline
		$\bullet$ Distinguishing registered rescue UAVs from unknown aircraft.
		\\
		\midrule
		Low-altitude agriculture
		&
		$\bullet$ Uplink rate $>25$ Mbps for high-definition video. \newline
		$\bullet$ End-to-end latency $<100$ ms.
		&
		$\bullet$ Sub-meter positioning accuracy along planned routes. \newline
		$\bullet$ Detection of small dynamic targets and ambient weather conditions.
		&
		$\bullet$ Continuous recognition of authorized UAVs. \newline
		$\bullet$ Support for regulatory traceability.
		\\
		\bottomrule
	\end{tabular}
\end{table*}

\subsubsection{Low-altitude smart logistics}
Low-altitude smart logistics leverages highly agile UAV swarms to form 3D point-to-point delivery networks that address the congestion bottlenecks of traditional ground transportation~\cite{re:lae-application-2025-2}. The primary communication demand stems from the need for continuous flight control and real-time route updates, while payload data including onboard camera feeds for route verification impose additional uplink requirements. 
In urban environments, densely distributed obstacles require continuous sensing for collision avoidance, and ground infrastructure must maintain persistent tracking of all active UAVs to support traffic management. 
From an identification perspective, the coexistence of authorized delivery UAVs and potential unauthorized intruders in the same airspace makes real-time identity acquisition a fundamental safety requirement.

\subsubsection{Urban air mobility}
UAM represents a paradigm shift in urban transportation, enabling point-to-point aerial travel within and between cities using eVTOL aircraft~\cite{re:lae-application-2021-3}. UAM operations are characterized by high passenger safety requirements, dense urban airspace, and frequent takeoff and landing cycles at vertiports distributed throughout the city. 
The safety-critical nature of passenger transport imposes some of the most stringent communication and sensing requirements among LAE scenarios, as any failure in flight control, obstacle detection, identity acquisition, or authentication can have severe consequences.
In particular, the simultaneous operation of multiple eVTOL aircraft in shared urban corridors demands precise spatial awareness and unambiguous identification of every airspace participant at all times.

\subsubsection{Low-altitude emergency rescue}
In disaster response scenarios where ground infrastructure has been damaged or destroyed, low-altitude aircraft provide rapidly deployable platforms for search and rescue, damage assessment, and emergency material delivery~\cite{re:lae-application-2025-4}. Operations take place in highly dynamic environments with critical timing constraints and severely degraded communication conditions. The key challenge is maintaining reliable connectivity and sensing capability in the absence of normal infrastructure, while accurately distinguishing rescue platforms from potentially hostile or malfunctioning aircraft under conditions where cooperative identification protocols may be unavailable.

\subsubsection{Low-altitude agriculture}
Low-altitude aircraft are increasingly deployed in precision agriculture for crop monitoring, pesticide spraying, seeding, and fertilization~\cite{re:lae-application-2025-5}. 
UAVs conduct high-resolution multispectral surveys to assess plant health and detect anomalies, while also executing precise chemical or seed dispensing over target areas. 
The dominant communication demand arises from high-definition video uplink for crop monitoring, and sensing must support accurate route execution to ensure complete field coverage. 
Identification in this scenario focuses on confirming the registered identity of operating UAVs and maintaining regulatory traceability throughout the mission.

\subsection{Summary}
This section reviewed the standalone approaches to low-altitude communication, sensing, and identification in LAINs, as well as several representative application scenarios. 
The above discussions indicate that these three functions are not isolated in practical low-altitude operations, but instead exhibit strong coupling in terms of service requirements, operational objectives, and airspace management needs. 
This motivates the transition from standalone functional support toward more integrated design. 
As the first step in this evolution, the next section reviews low-altitude ISAC, in which communication and sensing are jointly designed.

\section{ISAC for LAINs}
\label{sec:low-isac}

ISAC has emerged as a key paradigm for LAINs, allowing LAINs to support reliable data transmission for aerial and ground users, while achieving target localization, velocity estimation, and trajectory tracking through reflected signals~\cite{re:lae-networks-2026-6, re:lae-magazine-2025-isac-5, re:lae-magazine-2025-isac-4, re:lae-magazine-2025-isac-3, re:lae-isac-2026-3, re:lae-clutter-2026-1, re:lae-isac-2026-1}. 
In this section, we review the fundamentals of low-altitude ISAC, including system architectures, channel modeling, unified ISAC signal models, and performance metrics.
Then, we discuss communication-centric and sensing-centric waveform design and corresponding receiver design in detail, and further explore resource allocation and aircraft trajectory optimization in low-altitude ISAC.

\subsection{Fundamentals of low-altitude ISAC}

\subsubsection{System architectures}
\begin{figure*}[!t]
	\centering
	\includegraphics[width=0.95\textwidth]{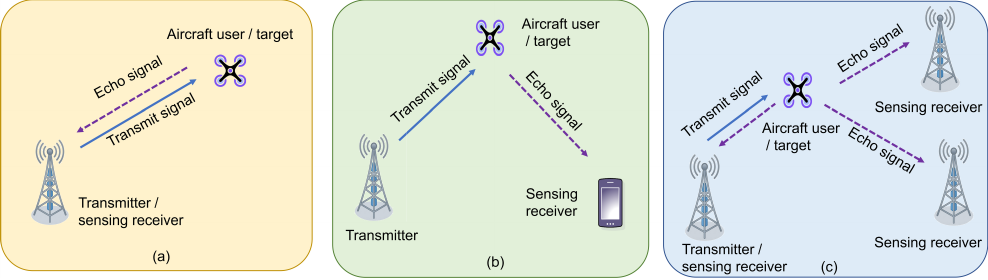}	
	\caption{Typical low-altitude ISAC deployment architectures: (a) Monostatic architecture with co-located transmitter and sensing receiver; (b) Bistatic architecture with separated transmitter and sensing receiver; (c) Distributed architecture with multiple cooperating transmitters and receivers.}
	\label{fig:isac_arc}
\end{figure*}

As shown in Figure~\ref{fig:isac_arc}, low-altitude ISAC systems can be classified by the physical deployment of their sensing transceivers into monostatic, bistatic, and distributed architectures~\cite{re:lae-future-waveform-2022-1}.
\begin{itemize}
	\item
	\textbf{Monostatic low-altitude ISAC:}
	The ISAC transmitter and sensing receiver are co-located to support downlink communication and echo reception~\cite{re:lae-isac-spc-2025-2}.
	While simple to implement, this architecture suffers from severe self-interference between the high-power transmitted signals and weak target echoes, and single-perspective observation limits spatial resolution and geometric diversity.
	
	\item
	\textbf{Bistatic low-altitude ISAC:}
	The transmitter and sensing receiver are spatially separated, where a base station handles downlink communication and signal illumination while an independent node captures aircraft echoes.
	The spatial separation avoids the self-interference of co-located full-duplex operation and introduces angular diversity for 3D target localization, at the cost of strict inter-station time and phase synchronization and reliable backhaul links.
	
	\item
	\textbf{Distributed low-altitude ISAC:}   
	Multiple base stations form a virtual large-aperture network, where each node independently receives echoes from the same aircraft and forwards measurements to a central fusion node for joint processing.
	The resulting geometric diversity enables high-precision 3D localization through time difference of arrival or AoA, and overcomes the elevation resolution limitations of single-station arrays.
	Since aircraft are highly likely to lie within the LoS coverage of multiple base stations~\cite{re:3gpp_tr36777}, distributed ISAC is particularly well suited for low-altitude scenarios, despite the challenges of strict synchronization and heavy backhaul overhead~\cite{re:lae-multipoint-2025-12}.
\end{itemize}

\subsubsection{Channel modeling}
Channel modeling is the prerequisite for the system design and performance analysis of low-altitude ISAC~\cite{re:lae-isac-channel-survey-2025}. 
This subsection focuses on monostatic low-altitude ISAC systems; similar modeling methods apply to bistatic and distributed systems.
Since aircraft operate in 3D space, the ISAC base station is assumed to utilize uniform planar arrays (UPAs) to transmit downlink ISAC signals and receive echo signals.
The steering vector of an $M_x \times M_z$ UPA for a target with an azimuth angle $\theta$ and an elevation angle $\phi$ is expressed as~\cite{van2002optimum} $\mathbf{a}(\theta,\phi) = \mathbf{a}_x(\theta,\phi) \otimes \mathbf{a}_z(\phi)$, where $\mathbf{a}_x(\theta,\phi) = \left[1, e^{j \frac{2\pi}{\lambda} d_x \sin\theta \cos\phi}, \cdots, e^{j \frac{2\pi}{\lambda}(M_x-1)d_x \sin\theta \cos\phi} \right]^T$ and $\mathbf{a}_z(\phi) = \left[1, e^{j \frac{2\pi}{\lambda} d_z \sin\phi}, \cdots, e^{j \frac{2\pi}{\lambda}(M_z-1)d_z \sin\phi} \right]^T$; $d_x$ and $d_z$ represent the horizontal and vertical antenna element spacings, respectively; $\lambda$ denotes the wavelength.

Based on the above array configuration, the channel models for monostatic low-altitude ISAC are characterized as follows.
\begin{itemize}
	\item
	\textbf{One-way communication channel:}
	Since the user (e.g., aircraft) is equipped with a single receive antenna, the one-way communication channel can be expressed as
	\begin{align}
		\mathbf{h}^{\mathrm{com}}(\tau,t) = \sum_{l=0}^{L-1} h_l^{\mathrm{com}}\, \delta(\tau-\tau_l^{\mathrm{com}})\, e^{j2\pi\nu_l^{\mathrm{com}} t}\, \mathbf{a}^*_t(\theta_l^{\mathrm{com}},\phi_l^{\mathrm{com}}),
		\label{eq:3-1-2-1}
	\end{align}
	where $l=0$ denotes the LoS path and $l \geq 1$ denotes the non-line-of-sight paths; $h_l^{\mathrm{com}}$ represents the effective path gain of the $l$-th path; $\tau_l^{\mathrm{com}}$ and $\nu_l^{\mathrm{com}}$ represent the path delay and Doppler frequency shift, respectively; $(\theta_l^{\mathrm{com}}, \phi_l^{\mathrm{com}})$ represent the azimuth and elevation AoDs of the $l$-th communication path at the base station, respectively.
	
	\item
	\textbf{Two-way sensing channel:}
	Assuming $Q$ targets exist in the airspace, the two-way sensing channel is expressed as
	\begin{align}
		\mathbf{H}^{\mathrm{sen}}(\tau,t) = \sum_{q=0}^{Q-1} h_q^{\mathrm{sen}} \sqrt{\sigma_q^{\mathrm{sen}}}\, \delta\!\left(\tau - \tau_q^{\mathrm{sen}}\right) e^{j2\pi \nu_q^{\mathrm{sen}} t}\, \mathbf{a}_r(\theta_q^{\mathrm{sen}},\phi_q^{\mathrm{sen}})\mathbf{a}_t^H(\theta_q^{\mathrm{sen}},\phi_q^{\mathrm{sen}}),
		\label{eq:3-1-2-2}
	\end{align}
	where $h_q^{\mathrm{sen}}$ represents the two-way path gain of the $q$-th target, and $\sigma_q^{\mathrm{sen}}$ represents the target RCS; $\tau_q^{\mathrm{sen}}$ and $\nu_q^{\mathrm{sen}}$ are the two-way delay and Doppler frequency shift, respectively; $\theta_q^{\mathrm{sen}}$ and $\phi_q^{\mathrm{sen}}$ denote the azimuth and elevation angles, respectively; $\mathbf{a}_t(\theta_q^{\mathrm{sen}},\phi_q^{\mathrm{sen}})$ and $\mathbf{a}_r(\theta_q^{\mathrm{sen}},\phi_q^{\mathrm{sen}})$ are the transmit and receive steering vectors, respectively.
	
	\item
	\textbf{RCS modeling:}
	RCS is a key parameter in sensing channels. RCS modeling methods can be broadly categorized into deterministic and statistical approaches. Deterministic methods, such as physical optics and geometrical optics, provide physics-based approximations of target scattering characteristics~\cite{re:knott2004radar}. Statistical methods characterize random RCS fluctuations using probability density functions. The Swerling I/II models assume exponentially distributed RCS with slow and fast fluctuations, respectively, while the Swerling III/IV models correspond to chi-square distributed RCS~\cite{re:swerling}. 
	In addition, to further improve RCS modeling precision for low-altitude targets,~\cite{re:lae-isac-channel-2026-1} proposes a unified RCS modeling framework that decomposes $\sigma_q^{\mathrm{sen}}$ into three components, namely large-scale power factors, small-scale angle-dependent components, and random components characterizing variations among target instances.
	
	\item
	\textbf{Clutter modeling:}
	The ISAC base station experiences severe clutter interference from environmental scatterers, comprising strong static clutter originating from ground surfaces and high-rise buildings alongside dynamic clutter generated by birds. Similar to \eqref{eq:3-1-2-2}, the clutter channel can be expressed as
	\begin{align}
		\mathbf{H}^{\mathrm{clu}}(\tau,t) = \sum_{i \in \mathcal{C}} h_i^{\mathrm{clu}} \sqrt{\sigma_i^{\mathrm{clu}}}\, \delta(\tau-\tau_i^{\mathrm{clu}})\, e^{j2\pi\nu_i^{\mathrm{clu}} t}\, \mathbf{a}_r(\theta_i^{\mathrm{clu}},\phi_i^{\mathrm{clu}})\mathbf{a}_t^H(\theta_i^{\mathrm{clu}},\phi_i^{\mathrm{clu}}),
		\label{eq:3-1-2-3}
	\end{align}
	where $h_i^{\mathrm{clu}}$ is the path gain of the $i$-th clutter scatterer; $\sigma_i^{\mathrm{clu}}$ is its clutter RCS; $\tau_i^{\mathrm{clu}}$ and $\nu_i^{\mathrm{clu}}$ denote the two-way delay and Doppler frequency shift; $\mathcal{C}$ denotes the index set of clutter scatterers; $\theta_i^{\mathrm{clu}}$ and $\phi_i^{\mathrm{clu}}$ denote the azimuth and elevation angles, respectively. 
\end{itemize}

\textbf{Existing works:} In addition to the aforementioned classical low-altitude ISAC channel models, a geometry-based stochastic model (GBSM) for UAV-assisted low-altitude ISAC systems is proposed in~\cite{re:lae-isac-channel-2025-3}, which fully considers UAV flight attitudes, movement trajectories, and scatterer distributions to capture the non-stationary characteristics of low-altitude environments.
The latest progress for 3GPP Release 19 ISAC channel standardization is comprehensively reviewed in~\cite{re:lae-isac-channel-2025-5}, which covers low-altitude ISAC channel modeling and also proposes an extended-GBSM integrating multiple key features such as RCS, monostatic background channels, micro-Doppler effects, and shared clusters.
Furthermore, a novel object-oriented low-altitude ISAC channel modeling framework is proposed in~\cite{re:lae-isac-channel-2026-2}, which jointly utilizes UAV-based measurements and ground measurements alongside ray-tracing technologies to achieve the modeling of both static background channels and dynamic channels.
These channel models primarily differ in the tradeoff between analytical tractability and environmental fidelity. 
For example, classical parametric models are analytically tractable and suitable for system-level performance analysis, but may not fully capture the complex scattering characteristics of low-altitude environments.
GBSMs provide a more flexible statistical description of channel non-stationarity, spatial consistency, and shared scattering clusters, but require additional model parameters and scenario-dependent calibration.

\subsubsection{Unified ISAC signal models}
This part provides unified ISAC signal models based on the channel models established in Section 3.1.2. 

\begin{itemize}
	\item 
	\textbf{Transmitted ISAC signal model:}
	For a single transmitted stream with a fixed beamforming vector, the ISAC signal is expressed as
	\begin{align}
		\mathbf{s}(t) = \mathbf{w}x(t),
		\label{eq:3-1-3-1}
	\end{align}
	where $x(t)$ denotes the transmitted signal in the time domain, and $\mathbf{w}$ represents the beamforming vector. This representation makes no assumptions about the underlying modulation waveform, and therefore OFDM, orthogonal time frequency space (OTFS), and FMCW can all be incorporated into this framework.
	
	\item
	\textbf{Received signal model at the communication receiver:}
	According to the communication channel $\mathbf{h}^{\mathrm{com}}(\tau, t)$ in \eqref{eq:3-1-2-1}, the received signal at the communication receiver is given by
	\begin{align}
		r^{\mathrm{com}}(t) =  
		\sum_{l=0}^{L-1}h_l^{\mathrm{com}}\mathbf{a}_t^H(\theta_l^{\mathrm{com}}, \phi_l^{\mathrm{com}})\mathbf{w}x(t-\tau_l^{\mathrm{com}})e^{j2\pi\nu_l^{\mathrm{com}} t} + z^{\mathrm{com}}(t),
		\label{eq:3-1-3-2}
	\end{align}
	where $z^{\mathrm{com}}(t)$ represents the additive white Gaussian noise (AWGN).
	
	\item
	\textbf{Received signal model at the sensing receiver:}
	According to the sensing channel $\mathbf{H}^{\mathrm{sen}}(\tau, t)$ in \eqref{eq:3-1-2-2} and clutter channel $\mathbf{H}^{\mathrm{clu}}(\tau, t)$ in \eqref{eq:3-1-2-3}, and assuming that the self-interference has been effectively suppressed, the received signal at the sensing receiver is expressed as
	\begin{align}
		\mathbf{r}^{\mathrm{sen}}(t) ={} &\underbrace{\sum_{q=0}^{Q-1} h_q^{\mathrm{sen}} \sqrt{\sigma_q^{\mathrm{sen}}}\,\mathbf{a}_r(\theta_q^{\mathrm{sen}},\phi_q^{\mathrm{sen}})\mathbf{a}_t^H(\theta_q^{\mathrm{sen}},\phi_q^{\mathrm{sen}})\mathbf{w}x\!\left(t - \tau_q^{\mathrm{sen}}\right)e^{j2\pi\nu_q^{\mathrm{sen}}t}}_{\text{target echoes}} 
		\notag \\
		&+ \underbrace{\sum_{i\in\mathcal{C}}h_i^{\mathrm{clu}} \sqrt{\sigma_i^{\mathrm{clu}}}\,\mathbf{a}_r(\theta_i^{\mathrm{clu}},\phi_i^{\mathrm{clu}})\mathbf{a}_t^H(\theta_i^{\mathrm{clu}},\phi_i^{\mathrm{clu}})\mathbf{w}x(t-\tau_i^{\mathrm{clu}})e^{j2\pi\nu_i^{\mathrm{clu}} t}}_{\text{clutter}} + \mathbf{z}^{\mathrm{sen}}(t),
		\label{eq:3-1-3-3}
	\end{align}
	where $\mathbf{z}^{\mathrm{sen}}(t)$ denotes the AWGN vector.
\end{itemize}

\subsubsection{Performance metrics}
ISAC systems require performance evaluation from three dimensions: communication, sensing, and their joint tradeoff~\cite{re:lae-future-waveform-2022-1}.

\begin{itemize}
	\item
	\textbf{Communication performance metrics:}
	Communication performance of ISAC can be evaluated in terms of efficiency and reliability~\cite{re:goldsmith2005wireless}. Efficiency is primarily measured by the achievable rate, while channel capacity provides a theoretical upper bound. Other efficiency-related metrics include coverage and user capacity. 
	Reliability is mainly characterized by the bit error rate (BER), which is particularly critical for mission-critical command-and-control links, and the outage probability, which captures link robustness under channel fluctuations. 
	End-to-end delay also constitutes a key constraint in real-time UAV control scenarios.

	\item
	\textbf{Sensing performance metrics:}
	Sensing performance of ISAC is evaluated from two aspects: detection and estimation. Detection performance is characterized by the receiver operating characteristic curve, which describes the tradeoff between detection probability $P_d$ and false alarm probability $P_{fa}$ at a given sensing signal-to-interference-plus-noise ratio (SINR)~\cite{re:lae-radar-2005-1}.
	Estimation performance can be measured by the mean square error (MSE) and Cram\'{e}r--Rao lower bound (CRLB)~\cite{re:kay1993statistical}, where the CRLB serves as the theoretical lower bound on the variance of any unbiased estimator and is derived by inverting the Fisher information matrix (FIM). Range resolution and velocity resolution are complementary system-level descriptors that determine whether adjacent targets can be resolved.

	\item
	\textbf{Joint performance metrics:}
	ISAC systems inherently face a tradeoff between communication and sensing under shared power and bandwidth constraints. This tradeoff can be characterized by the Pareto frontier of the achievable communication and sensing performance. Weighted-sum optimization with a weight factor $\rho\in[0,1]$ is commonly used to explore different communication--sensing operating points~\cite{re:on-the-funda-fanliu-2023}.
\end{itemize}

\subsection{Communication-centric ISAC}
Communication-centric ISAC aims to utilize or adapt existing communication waveforms to achieve target sensing. 
These waveforms prioritize communication performance, and they exhibit high compatibility with existing communication standards and infrastructure. Consequently, communication-centric ISAC is particularly suitable for low-altitude application scenarios where communication quality of service constitutes the primary requirement.

OFDM has been widely adopted as the standard physical-layer waveform for 4G and 5G, and current 3GPP 6G discussions also support cyclic prefix (CP)-OFDM as a baseline candidate for the downlink waveform.
It is also the most widely studied and validated
communication-centric ISAC waveform in current ISAC research.
Therefore, this section focuses primarily on OFDM-enabled low-altitude ISAC systems by reviewing existing clutter suppression, parameter estimation, and data demodulation methods, followed by a discussion of novel waveforms tailored for low-altitude high-mobility scenarios, such as OTFS~\cite{re:lae-otfs-2022-1} and affine frequency division multiplexing (AFDM)~\cite{re:lae-isac-waveform-2026-3}.

\subsubsection{Transmitted OFDM signal}
Let $M$, $N$, $\Delta f$, $T = 1/\Delta f$, $T_{\mathrm{cp}}$, $T_{\mathrm{sym}} = T + T_{\mathrm{cp}}$, $B = M\Delta f$, and $x_{m,n}$ respectively denote the number of subcarriers, number of OFDM symbols, subcarrier spacing, symbol duration without CP, CP duration, total symbol duration with CP, system bandwidth, and the frequency-domain transmitted symbol on the $m$-th subcarrier in the $n$-th OFDM symbol. Then, according to Section 3.1.3, the transmitted OFDM signal is expressed as
\begin{align}
	\mathbf{s}(t) = \sum_{n=0}^{N-1}\sum_{m=0}^{M-1}\mathbf{w}\, x_{m,n}\, e^{j2\pi m\Delta f(t - nT_{\mathrm{sym}} - T_{\mathrm{cp}})}\,\mathrm{rect}\!\left(\frac{t - nT_{\mathrm{sym}}}{T_{\mathrm{sym}}}\right),
	\label{eq:3-2-1-1}
\end{align}
where $\mathrm{rect}(t)=1$ for $t \in [0,1)$, and $\mathrm{rect}(t)=0$ otherwise.

\subsubsection{Parameter estimation at the sensing receiver}
According to \eqref{eq:3-1-3-3} in Section~3.1.3 and assuming that the CP duration exceeds the maximum two-way delay, the frequency-domain signal at the $m$-th subcarrier of the $n$-th OFDM symbol received by the sensing receiver is expressed as
\begin{align}
	\mathbf{y}_{m,n}^{\mathrm{sen}} = &\underbrace{x_{m,n}\sum_{q=0}^{Q-1}\tilde{h}_q^{\mathrm{sen}}\,\mathbf{a}_r(\theta_q^{\mathrm{sen}},\phi_q^{\mathrm{sen}})\, e^{-j2\pi m\Delta f\tau_q^{\mathrm{sen}}}\, e^{j2\pi nT_{\mathrm{sym}}\nu_q^{\mathrm{sen}}}\,D_M(\epsilon_q^{\mathrm{sen}})}_{\text{desired term (including gain loss)}} \notag \\ &+ \underbrace{\sum_{q=0}^{Q-1}\mathbf{a}_r(\theta_q^{\mathrm{sen}},\phi_q^{\mathrm{sen}})\sum_{m'\neq m}\tilde{h}_q^{\mathrm{sen}}\,x_{m',n}\,e^{-j2\pi m'\Delta f\tau_q^{\mathrm{sen}}}\,e^{j2\pi nT_{\mathrm{sym}}\nu_q^{\mathrm{sen}}}\,D_M(m'-m+\epsilon_q^{\mathrm{sen}})}_{\text{ICI term}} + \mathbf{z}_{m,n}^{\mathrm{sen}},
	\label{eq:3-2-2-1}
\end{align}
where $\tilde{h}_q^{\mathrm{sen}}=h_q^{\mathrm{sen}}\sqrt{\sigma_q^{\mathrm{sen}}}\,\mathbf{a}_t^H(\theta_q^{\mathrm{sen}},\phi_q^{\mathrm{sen}})\mathbf{w}\cdot e^{j2\pi\nu_q^{\mathrm{sen}} T_{\mathrm{cp}}}$ is the effective gain; $\epsilon_q^{\mathrm{sen}} = \nu_q^{\mathrm{sen}}/\Delta f$ is the normalized Doppler shift; $D_M(\epsilon) = \frac{1}{M}\cdot\frac{1 - e^{j2\pi\epsilon}}{1 - e^{j2\pi\epsilon/M}}$ is the $M$-point Dirichlet kernel; $\mathbf{z}_{m,n}^{\mathrm{sen}}$ contains the clutter interference and AWGN.
Aircraft movements in low-altitude environments cause $\epsilon_q^{\mathrm{sen}}$ to become non-zero, consequently generating ICI. 
Furthermore, a third-order tensor $\boldsymbol{\mathcal{Y}}$ is constructed, where $\boldsymbol{\mathcal{Y}}(:,m,n)=\mathbf{y}_{m,n}^{\mathrm{sen}}$.
The sensing receiver needs to estimate aircraft physical parameters $\{\theta_q^{\mathrm{sen}},\phi_q^{\mathrm{sen}},\tau_q^{\mathrm{sen}},\nu_q^{\mathrm{sen}}\}_{q=0}^{Q-1}$ from $\boldsymbol{\mathcal{Y}}$ to subsequently achieve 3D localization and velocity estimation for aircraft.

\textbf{Mitigating clutter is essential for reliable parameter estimation}~\cite{re:lae-clutter-2026}, as strong static clutter and dynamic clutter overlap in low-altitude environments and severely degrade parameter estimation performance.
Existing clutter suppression works for OFDM-enabled low-altitude ISAC are broadly divided into two categories. 
\begin{itemize}
	\item
	The first category comprises traditional methods based on mathematical models, where model-driven methods exploit explicit structural or statistical knowledge of clutter.
	For example, moving target indication (MTI) utilizes the characteristic that static clutter possesses near-zero Doppler frequency shifts to separate clutter from moving target echoes~\cite{re:lae-clutter-2024-7}, whereas channel knowledge map-assisted space--time adaptive processing further uses stored angular information to suppress clutter that overlaps with low-speed targets in the Doppler domain~\cite{re:lae-ckm-2025-7}.
	When the clutter statistics are unavailable or uncertain, the clutter covariance matrix can instead be treated as a latent variable and jointly inferred with the target parameters through variational Bayesian learning~\cite{re:lae-ofdm-clutter-2025-1}.
	Robust transmit--receive design provides another route by forming spatial nulls toward clutter, where Bernstein inequalities are utilized to handle clutter parameter estimation errors to achieve robust clutter suppression~\cite{re:lae-ofdm-clutter-2025-2}.
	
	\item
	The second category is AI-enabled clutter suppression, where AI methods such as machine learning and deep learning significantly enhance the robustness of clutter suppression through data-driven adaptive learning and multi-dimensional feature mining~\cite{re:lae-clutter-2026}. 
	In~\cite{re:lae-clutter-2026-1}, clutter is regarded as normal samples and target echoes are treated as abnormal data based on practical OFDM-enabled ISAC system deployments. 
	This work successfully achieves the detection of weak low-speed UAVs at a distance of one kilometer.
	A recurrent neural network incorporating long short-term memory (LSTM) and attention mechanisms is proposed in~\cite{re:lae-ofdm-clutter-2025-4}, where time--frequency features are extracted on a per-OFDM-symbol basis and outlier errors induced by clutter are suppressed. 
	Robust parameter estimation is successfully achieved without requiring explicit clutter suppression.
\end{itemize}

The above clutter-suppression methods exhibit different tradeoffs among model dependence, environmental adaptability, and implementation complexity.
For example, AI-enabled methods can learn complicated clutter characteristics without an explicit parametric model, while their robustness across deployment sites and operating conditions depends on the representativeness of the training data.
Therefore, practical method selection needs to balance low-speed target preservation, prior-information requirements, computational complexity, and adaptability to changing clutter environments.

\textbf{Furthermore, high-precision and low-complexity parameter estimation methods are of paramount importance.}
When aircraft fly at low speeds or remain in hovering states, $|\nu_q^{\mathrm{sen}}| \ll \Delta f$ and $D_M(\epsilon_q^{\mathrm{sen}}) \approx 1$, and thus the effect of ICI becomes negligible. 
In this case, after the known non-zero data symbols are removed, the noiseless response of each target is approximately separable across the spatial, subcarrier, and slow-time dimensions.
Therefore, classical estimation algorithms, such as periodogram-based methods~\cite{re:lae-isac-spc-2024-4}, subspace decomposition-based methods~\cite{re:lae-isac-spc-2021-5}, and compressive sensing (CS)-based methods~\cite{re:lae-isac-spc-2026-6}, can be directly applied to accomplish parameter estimation tasks.
In~\cite{re:lae-ofdm-clutter-2026-5}, AoAs are estimated utilizing the multiple signal classification (MUSIC) algorithm, and delays and Doppler frequencies are jointly estimated utilizing a 2D periodogram, which successfully achieves robust parameter estimation for multiple UAVs.
In~\cite{re:lae-multipoint-2025-4,re:lae-multipoint-2025-11}, CANDECOMP/PARAFAC decomposition-based methods are introduced for collaborative aircraft sensing by multiple base stations. 
These methods exploit the inherent Vandermonde structure of the factor matrices to estimate AoAs, delays, and Doppler frequencies.
Furthermore, the theoretical relationship between base station sensing channels and UAV six-dimensional motion parameters is derived in~\cite{re:lae-isac-spc-2025-2}, while in~\cite{re:lae-multipoint-2026-10} a physics-embedded learning method is proposed to mitigate off-grid errors in traditional CS-based methods.

ICI becomes non-negligible when aircraft move at high speeds or high-frequency signals like millimeter-wave are adopted. 
For instance, when $|\epsilon_q^{\mathrm{sen}}|=0.5$ and $M$ is sufficiently large, the desired-subcarrier power is approximately 40\% of the total, while roughly 60\% of the signal power leaks into other subcarriers as ICI.
To address this issue, an ICI-aware parameter estimation framework is proposed in~\cite{re:lae-isac-spc-2021-7}. 
Delay--Doppler estimation is modeled as a joint carrier frequency offset and channel estimation problem, and joint angle--delay--Doppler estimation for multiple targets is achieved utilizing amplitude and phase estimation spatial filtering alongside an orthogonal matching pursuit (OMP)-based iterative interference cancellation mechanism. 
A general and structured fast/slow-time representation is first derived in~\cite{re:lae-isac-waveform-2025-5}, based on which an ambiguity function-assisted active sensing scheme is proposed. 
It consists of three modules, i.e., a coarse search and triangular ambiguity function matching module, a gradient-based fine-tuning module, and a modified cell-averaging constant false alarm rate (CA-CFAR) target detection module.
The reviewed parameter-estimation methods reveal a further tradeoff among estimation resolution, high-mobility robustness, signaling overhead, and receiver complexity.
For example, conventional periodogram-based and other separable estimators are computationally efficient, but their performance can degrade rapidly when ICI becomes significant.
ICI-aware parameter estimation methods can maintain high estimation accuracy in high-mobility scenarios, at the cost of increased receiver complexity.

\subsubsection{Data demodulation at the communication receiver}
According to \eqref{eq:3-1-3-2} in Section~3.1.3, the frequency-domain signal at the $m$-th subcarrier of the $n$-th OFDM symbol received by the communication user is expressed as
\begin{align}
	y_{m,n}^{\mathrm{com}} = \sum_{l=0}^{L-1} \sum_{m'=0}^{M-1} \tilde{h}_l^{\mathrm{com}} x_{m',n}e^{j2\pi\nu_l^{\mathrm{com}} nT_{\mathrm{sym}}}e^{-j2\pi m'\Delta f\tau_l^{\mathrm{com}}} \cdot D_M(m'-m+\epsilon_l^{\mathrm{com}}) + z_{m,n}^{\mathrm{com}},
	\label{eq:3-2-3-1}
\end{align}
where $\tilde{h}_l^{\mathrm{com}} = h_l^{\mathrm{com}}\,\mathbf{a}_t^H(\theta_l^{\mathrm{com}},\phi_l^{\mathrm{com}})\mathbf{w}\cdot e^{j2\pi\nu_l^{\mathrm{com}} T_{\mathrm{cp}}}$; $\epsilon_l^{\mathrm{com}} = \nu_l^{\mathrm{com}}/\Delta f$ is the normalized Doppler shift of the $l$-th communication path; $z_{m,n}^{\mathrm{com}}$ denotes the corresponding AWGN. 
Then, the communication receiver needs to demodulate data $\{x_{m,n}\}$ from $\{y_{m,n}^{\mathrm{com}}\}$.

When $|\epsilon_l^{\mathrm{com}}| \ll 1$, the ICI term becomes negligible. 
The communication channel reduces to an approximate single-tap frequency-domain model. 
Traditional channel-estimation methods, including least squares (LS) and minimum mean square error (MMSE) estimation, can be directly applied~\cite{re:lae-ofdm-ce-1998-1,re:lae-ofdm-ce-2002-2}.
Furthermore, traditional data detection methods including zero-forcing equalization and MMSE equalization are directly applied at the communication receiver. 
When ICI is non-negligible, specialized signal processing techniques are required. 
Regarding channel estimation, low-altitude communication channels are dominated by strong LoS components with a limited number of paths. 
Since these channels exhibit significant sparsity in the delay--Doppler domain~\cite{re:cellular-uav-survey-2019,re:uav-com-2019}, CS-based methods are utilized to perform sparse reconstruction of communication channels by exploiting this inherent sparse structure~\cite{re:lae-ofdm-ce-2016-3}. 
Methods including OMP and sparse Bayesian learning (SBL) accurately recover sparse channel coefficients in the delay--Doppler domain utilizing only a small number of pilot observations. 
Regarding data detection, non-negligible ICI causes the frequency-domain channel matrix to exhibit a banded off-diagonal structure. 
This can be processed utilizing iterative interference cancellation receivers, factor graph-based message passing algorithms like belief propagation, or turbo iterative receivers that perform joint equalization and decoding~\cite{re:lae-ofdm-ce-2004-4}. These methods progressively suppress residual ICI, and reliably recover the transmitted data.
These receiver methods exhibit a clear tradeoff between implementation simplicity and robustness to Doppler-induced ICI.
Therefore, the receiver architecture should be selected according to aircraft mobility, channel sparsity, available pilot resources, and real-time processing constraints.

\subsubsection{New communication-centric ISAC waveforms for high-mobility scenarios}
Although OFDM provides high compatibility with existing communication standards, its subcarrier orthogonality becomes difficult to maintain in highly time-varying low-altitude channels.
Thus, novel waveforms such as OTFS and AFDM are also investigated for low-altitude high-mobility scenarios. 
\begin{itemize}
	\item 
	\textbf{OTFS:}
	OTFS performs signal processing in the delay--Doppler domain, which transforms fast time-varying channels into quasi-static and sparse channels. Consequently, it exhibits exceptional resistance against multipath and Doppler shift effects~\cite{re:lae-otfs-2024-2}.
	The mathematical foundation for achieving communication and radar sensing in the delay--Doppler domain utilizing OTFS is elaborated in~\cite{re:lae-otfs-2022-1}. 
	In~\cite{re:lae-otfs-2025-7}, relative echo delays and Doppler shifts extracted from OTFS channel estimation are utilized for 3D localization and velocity estimation of UAVs through iterative algorithms combining belief propagation and gradient descent. This work effectively validates the application potential of OTFS in low-altitude ISAC.
	Furthermore, OTFS-assisted ISAC research is systematically reviewed in~\cite{re:lae-otfs-2024-6}, where the robustness of OTFS in high-Doppler scenarios is highlighted through comparisons between OTFS-based and OFDM-based ISAC systems.    
	
	\item
	\textbf{AFDM:}
	AFDM multiplexes information symbols in the discrete affine Fourier transform (DAFT) domain using chirp basis functions for modulation, which, with appropriately chosen chirp parameters, allows AFDM to effectively overcome ICI and achieve full diversity gains in time- and frequency-selective channels~\cite{re:lae-isac-waveform-2026-6}.
	In~\cite{re:lae-isac-waveform-2026-3}, the CRLB for AFDM-based sensing and the ambiguity function of a pilot-assisted AFDM waveform are derived to identify conditions for stable sensing performance.
	The work of~\cite{re:lae-isac-waveform-2026-6} investigates the application of AFDM in secure low-altitude ISAC, and reveals the advantages of AFDM in achieving sensing-assisted secure transmission mechanisms within low-altitude scenarios.
\end{itemize}

Unlike OFDM, OTFS maps information symbols to the delay--Doppler domain, where the sparse and slowly varying channel representation facilitates the estimation of target delay and Doppler parameters, while the associated 2D modulation, channel estimation, and data detection increase receiver-processing overhead. 
AFDM employs chirp basis functions in the DAFT domain, offering another approach to high-mobility transmission with potentially lower channel-estimation and processing overhead than 2D OTFS.

\subsection{Sensing-centric ISAC}
Sensing-centric ISAC embeds communication data into radar waveforms while prioritizing sensing performance.
Compared to communication-centric ISAC in Section 3.2, it is more suitable for low-altitude application scenarios demanding high sensing precision.
Due to its strong resistance to Doppler shifts, extended sensing range, and simple self-interference suppression, FMCW has been widely recognized as a highly promising waveform for 6G sensing applications~\cite{re:lae-radar-2018-2,re:lae-fmcw-com-2022-5}. Therefore, this section focuses on FMCW-enabled low-altitude ISAC systems, covering the basic FMCW waveform, four schemes for embedding communication data into the FMCW waveform, and the corresponding sensing receiver design and communication receiver design.

\subsubsection{Basic FMCW waveform}
The ISAC base station continuously transmits $K$ chirps, and $T_c$ represents the single chirp duration. 
Then, the transmitted time-domain FMCW signal $x_b(t)$ is expressed as~\cite{re:fmcw-sa-2015,re:fmcw-ma-2023}
\begin{align}
	x_b(t) = \sum_{k=0}^{K-1} \exp\left(j2\pi\left(f_s (t - k T_c) + \frac{\mu}{2}(t - k T_c)^2\right)\right)\mathrm{rect}\left(\frac{t - k T_c}{T_c}\right),
	\label{eq:3-3-1-1}
\end{align}
where $f_s=f_c-B/2$ denotes the start frequency; $f_c$ is the carrier frequency; $\mu = B/T_c$ denotes the chirp slope; $B$ represents the sweep bandwidth. The constant-envelope nature of FMCW yields an approximately 0 dB peak-to-average power ratio (PAPR), enabling more efficient power amplifier operation than conventional high-PAPR OFDM waveforms~\cite{re:lae-radar-2018-2,re:lae-fmcw-com-2022-5}. Under a common peak-power constraint, this low-PAPR property allows FMCW to achieve a substantially higher average transmit power than OFDM, since the transmitter can be driven closer to saturation without clipping. The resulting increase in energy on target is particularly beneficial for detecting low-RCS objects such as small low-altitude UAVs.

\subsubsection{Schemes embedding communication data into FMCW}
Schemes embedding communication data into the FMCW waveform are broadly categorized into four types based on modulation dimensions.

\begin{itemize}
	\item \textbf{Inter-chirp phase modulation:} The information phase $\varphi_k$ is embedded into the $k$-th chirp, whose transmitted signal is~\cite{re:lae-fmcw-com-2022-5}
	\begin{equation}
		x_k(\hat{t}) =\exp\!\left\{j2\pi\!\left(f_s\hat{t} + \frac{\mu}{2}\hat{t}^2\right) + j\varphi_k\right\}  \mathrm{rect}\!\left(\frac{\hat{t}}{T_c}\right) ,
		\label{eq:3-3-2-1}
	\end{equation}
	where $\hat{t} = t - k T_c \in [0, T_c)$; $\varphi_k \in
	\left\{
	\frac{2\pi u}{M_{\mathrm{PSK}}}
	\,\middle|\,
	u=0,1,\ldots,M_{\mathrm{PSK}}-1
	\right\}$ corresponds to $M_{\mathrm{PSK}}$-ary phase shift keying (PSK) modulation.

	\item \textbf{Chirp-parameter modulation:} Structural parameters of chirps such as the slope $\mu$ or starting frequency are modulated to carry information~\cite{re:lae-fmcw-com-2022-5}. For example, for starting frequency shift keying (FSK), the transmitted signal is
	\begin{equation}
		x_k(\hat{t}) = \exp\!\left\{j2\pi\!\left(f_{s,d_k}\hat{t} + \frac{\mu}{2}\hat{t}^2\right)\right\}\mathrm{rect}\!\left(\frac{\hat{t}}{T_c}\right),
	\end{equation}
	where $f_{s,d_k}\in\{f_{s,0},f_{s,1},\ldots,f_{s,M_{\mathrm{FSK}}-1}\}$ is selected according to communication data $d_k$. 
	LoRa, the widely adopted low-power wide-area communication technology based on chirp spread spectrum modulation, can be viewed as a special case within chirp-parameter modulation.
	
	\item \textbf{Intra-chirp multiplicative modulation:} A communication baseband signal $d(\hat{t})$ is embedded into the FMCW chirp by multiplicative modulation~\cite{re:lae-fmcw-com-2021-1}
	\begin{equation}
		x_k(\hat{t}) = \exp\!\left\{j2\pi\!\left(f_s\hat{t}+\frac{\mu}{2}\hat{t}^2\right)\right\}\mathrm{rect}\!\left(\frac{\hat{t}}{T_c}\right) \cdot d(\hat{t}) .
	\end{equation}
	This scheme can achieve a high communication rate, but could disrupt the single-frequency structure after dechirp, introducing sidelobes and causing greater sensing performance degradation. 
	When $d(\hat{t})$ is a narrowband OFDM signal, the resulting waveform reduces to the OFDM-FMCW hybrid scheme.
	
	\item \textbf{Index modulation:} Information is carried by controlling chirp on-off statuses, subband selections, or transmit antennas~\cite{re:lae-fmcw-com-2021-4}. For chirp on-off index modulation, the transmitted signal is
	\begin{equation}
		x(t) = \sum_{k\in\mathcal{X}} \exp\!\left\{j2\pi\!\left(f_s(t-kT_c)+\frac{\mu}{2}(t-kT_c)^2\right)\right\}\mathrm{rect}\!\left(\frac{t-kT_c}{T_c}\right),
	\end{equation}
	where the active chirp index set $\mathcal{X}\subseteq\{0,1,\ldots,K{-}1\}$ carries the information. 
\end{itemize}

The four schemes differ in how strongly the communication symbols perturb the original sensing waveform.
For example, inter-chirp phase modulation preserves the fast-time structure and constant envelope of each chirp, but the data phase and the Doppler-induced phase rotation coexist in the slow-time dimension, thereby limiting straightforward data demodulation.
Chirp-parameter modulation retains the chirp structure, but varying the starting frequency or slope with the data introduces phase or beat-frequency variations across chirps, which must be compensated before target parameters are estimated.
In the following, we take the widely used inter-chirp phase modulation as an example to elaborate on parameter estimation and data demodulation~\cite{re:lae-fmcw-com-2022-5}.

\subsubsection{Parameter estimation at the sensing receiver}

According to \eqref{eq:3-1-3-3} in Section 3.1.3, after dechirp processing utilizing the transmitted signal in \eqref{eq:3-3-2-1}, the beat signal $\mathbf{y}_k^{\mathrm{sen}}(\hat{t})$ received by the sensing receiver at fast time $\hat{t}$ during the $k$-th chirp is expressed as
\begin{align}
	\mathbf{y}_k^{\mathrm{sen}}(\hat{t})= \sum_{q=0}^{Q-1} \tilde{h}_q^{\mathrm{sen}} e^{-j2\pi f_s \tau_q^{\mathrm{sen}}}\, \underbrace{\mathbf{a}_r(\theta_q^{\mathrm{sen}}, \phi_q^{\mathrm{sen}})}_{\text{spatial}}\underbrace{e^{j2\pi \nu_q^{\mathrm{sen}} k T_c}}_{\text{slow-time}} \underbrace{e^{-j2\pi\left(\mu \tau_q^{\mathrm{sen}} - \nu_q^{\mathrm{sen}}\right)\hat{t}}}_{\text{fast-time}} + \mathbf{z}_k^{\mathrm{sen}}(\hat{t}),
	\label{eq:3-3-3-1}
\end{align}
where $\tilde{h}_q^{\mathrm{sen}}=h_q^{\mathrm{sen}}\sqrt{\sigma_q^{\mathrm{sen}}}\,\mathbf{a}_t^H(\theta_q^{\mathrm{sen}},\phi_q^{\mathrm{sen}})\mathbf{w}$ is the effective gain; 
$\mathbf{z}_k^{\mathrm{sen}}(\hat{t})$ represents the corresponding superposition of noise and clutter. 
$\mathbf{y}_k^{\mathrm{sen}}(\hat{t})$ is then sampled in the fast-time dimension with sampling interval $T_s$, and a third-order tensor $\boldsymbol{\mathcal{Y}}$ is constructed with $\boldsymbol{\mathcal{Y}}(:,k,n_s) = \mathbf{y}_{k,n_s}^{\mathrm{sen}}=\mathbf{y}_k^{\mathrm{sen}}(\hat{t})|_{\hat{t}=n_sT_s}$.
The sensing receiver needs to estimate aircraft physical parameters $\{\theta_q^{\mathrm{sen}},\phi_q^{\mathrm{sen}},\tau_q^{\mathrm{sen}},\nu_q^{\mathrm{sen}}\}_{q=0}^{Q-1}$ from $\boldsymbol{\mathcal{Y}}$ to subsequently achieve 3D localization and velocity estimation for aircraft.

From \eqref{eq:3-3-3-1}, the transmitted communication symbols $\{\varphi_k\}_{k=0}^{K-1}$ are eliminated through dechirp processing, which indicates that \textbf{clutter suppression and parameter estimation schemes designed for conventional FMCW-based sensing systems are directly applicable at the sensing receiver.}

\begin{itemize}
	\item
	\textbf{Clutter suppression:}
	Existing methods range from low-complexity slow-time filtering to iterative signal decomposition and physics-informed learning.
	MTI filtering and clutter-map-based processing exploit the slow-time differences between clutter and moving targets, where MTI-like adjacent-chirp subtraction provides a low-complexity implementation for highlighting clutter-contaminated samples~\cite{re:lae-radar-2023-7}.
	These methods are attractive for real-time processing, although Doppler-based separation becomes more difficult when hovering or slowly moving aircraft exhibit Doppler characteristics close to those of stationary clutter.
	More sophisticated decomposition methods can better separate weak target components from strong and non-stationary clutter.
	For example, ensemble empirical mode decomposition combined with the whale optimization algorithm iteratively reconstructs signal subspaces and extracts individual target components~\cite{re:lae-radar-2023-6}.
	Such iterative decomposition and search improve processing flexibility, but introduce higher computational complexity than direct filtering.
	Physics-informed guided decomposition networks further combine explicit low-rank and sparse structures with learned sparsity-enhancement modules to preserve weak UAV features~\cite{re:lae-radar-2026-4}.
	This model-driven learning route reduces the dependence on fully specified clutter distributions, while requiring representative training data and additional neural-network inference.
	
	\item 
	\textbf{Parameter estimation:} 
As indicated by \eqref{eq:3-3-3-1}, the constructed tensor $\boldsymbol{\mathcal{Y}}$ is decoupled across the spatial, slow-time, and fast-time dimensions, reducing parameter estimation in each dimension to a spectral-estimation problem.
	Periodogram- and fast Fourier transform (FFT)-based methods have relatively low implementation complexity, but their resolution is constrained by the finite observation aperture and the discretized spectral grid~\cite{re:fmcw-fft}.
	Subspace-based methods, such as MUSIC, can provide finer parameter resolution, while requiring covariance estimation, subspace decomposition, and spectral search~\cite{re:fmcw-ma-2023}.
	CS-based methods exploit the sparse distribution of targets in the parameter domain, but generally involve iterative reconstruction and may be affected by off-grid mismatch~\cite{re:fmcw-omp}.
	A practical compromise is to combine a low-complexity coarse estimator with a high-resolution refinement stage. For example, the FMCW radar algorithm in~\cite{re:fmcw-fft} first obtains coarse range estimates via the FFT and then adaptively reduces the number of samples fed into MUSIC, achieving performance similar to that of conventional MUSIC with an average complexity reduction of 88\%.
	The applicability of these conventional estimators also depends on the selected data-embedding scheme.
	For chirp-parameter modulation, the data-induced beat-frequency offset must be compensated before parameter estimation, whereas intra-chirp multiplicative modulation requires additional sidelobe suppression or removal of the communication symbols before standard spectral estimation can be reliably applied~\cite{re:lae-fmcw-com-2022-5}.
	Therefore, the receiver design must jointly consider estimation resolution, computational complexity, and the distortion introduced by communication-data embedding.
\end{itemize}

\subsubsection{Data demodulation at the communication receiver}
According to \eqref{eq:3-1-3-2} in Section 3.1.3, after dechirp processing utilizing the basic FMCW signal in \eqref{eq:3-3-1-1}, the beat signal $y_k^{\mathrm{com}}(\hat{t})$ received by the communication user at fast time $\hat{t}$ during the $k$-th chirp is expressed as
\begin{align}
	{y}_k^{\mathrm{com}}(\hat{t})= \sum_{l=0}^{L-1} \tilde{h}_l^{\mathrm{com}} e^{-j2\pi f_s \tau_l^{\mathrm{com}}}\, \underbrace{e^{j2\pi \nu_l^{\mathrm{com}} k T_c}}_{\text{slow-time}} \underbrace{e^{-j2\pi\left(\mu \tau_l^{\mathrm{com}} -\nu_l^{\mathrm{com}}\right)\hat{t}}}_{\text{fast-time}} \underbrace{e^{j\varphi_k}}_{\text{data}} + {z}_k^{\mathrm{com}}(\hat{t}),
	\label{eq:3-3-4-1}
\end{align}
where $\tilde{h}_l^{\mathrm{com}} = h_l^{\mathrm{com}}\,\mathbf{a}_t^H(\theta_l^{\mathrm{com}},\phi_l^{\mathrm{com}})\mathbf{w}$, and $z_k^{\mathrm{com}}(\hat{t})$ denotes the corresponding AWGN. 
${y}_k^{\mathrm{com}}(\hat{t})$ is then sampled in the fast-time dimension with sampling interval $T_s$, and a matrix $\mathbf{Y}$ is constructed with $\mathbf{Y}(k,n_s) =y_{k,n_s}^{\mathrm{com}}=y_k^{\mathrm{com}}(\hat{t})|_{\hat{t}=n_sT_s}$.
Then, the communication receiver needs to demodulate data $\{\varphi_k\}$ from $\mathbf{Y}$.

From \eqref{eq:3-3-4-1}, after dechirp, the data term $e^{j\varphi_k}$ remains in the slow-time dimension and couples with the Doppler shift term.
Since both manifest as cross-chirp phase rotations, they cannot be directly separated: the Doppler-induced phase shift equally interferes with data demodulation, while the embedded data symbols broaden or even submerge the Doppler spectral peak, degrading Doppler frequency estimation accuracy.
To address this coupling issue, existing demodulation schemes are broadly categorized into three types~\cite{re:lae-fmcw-com-2022-5}. 
\begin{itemize}
	\item 
	The first is differential detection: by taking the phase difference between adjacent chirps, the Doppler contribution is reduced to a fixed bias, and the symbols are recovered via differential PSK detection without requiring explicit Doppler estimation, at the cost of an inherent BER penalty.
	
	\item 
	The second category relies on pilot symbols embedded within the communication data. 
	These pilots are utilized to obtain a coarse Doppler estimate, based on which the Doppler term is subsequently stripped from the slow-time sequence prior to data detection. 
	
	\item 
	The third category involves joint parameter estimation and communication data demodulation, where both components can be cooperatively resolved utilizing maximum likelihood (ML) or expectation-maximization algorithms.
\end{itemize}

The three demodulation strategies exhibit different tradeoffs among pilot overhead, detection reliability, and computational complexity.
For the remaining three data-embedding schemes, the demodulation approach differs accordingly. 
For example, for index modulation, the communication receiver determines the activation status of each chirp through energy detection or sparse recovery, and subsequently maps the detected activation pattern back to the transmitted data.

\subsection{Resource allocation and aircraft trajectory optimization}
Aircraft trajectory optimization and resource allocation, especially joint trajectory optimization and beamforming, are key technologies for low-altitude ISAC to cope with highly dynamic airspace environments and diverse service demands.
Compared to terrestrial ISAC, the controllable mobility of aircraft in low-altitude scenarios endows low-altitude ISAC systems with an additional optimization dimension.
Namely, the trajectory itself can participate in system optimization as a geometric resource.
Low-altitude ISAC actively adjusts the spatial positions of aircraft to change communication channel quality and sensing geometric conditions, thereby achieving a more flexible tradeoff between communication and sensing performance.
Existing low-altitude ISAC works on resource allocation and aircraft trajectory optimization can be broadly divided into two categories.

\begin{itemize}
	\item 
	\textbf{Aircraft acting as low-altitude communication users and targets:}
	Some works focus on UAVs acting as low-altitude communication users and targets, where terrestrial ISAC base stations provide communication services to UAVs while conducting airspace sensing.
	In~\cite{re:lae-isac-resource-2024-4}, a UAV and a base station form a bistatic synthetic aperture radar, where the UAV records the reflections of wireless downlink signals and sends them to the base station for object sensing and imaging.
	This study reveals a direct coupling between trajectory design, imaging quality, and energy consumption.
	In~\cite{re:lae-isac-resource-2026-3}, a deep reinforcement learning-based scheme is developed to jointly optimize base station beamforming and UAV trajectories, with the objective of maximizing the expected communication sum rate under the average sensing signal-to-noise ratio requirement, UAV flight missions, collision avoidance, and the maximum base station transmit power.
	The introduction of multiple cooperative base stations further provides spatial diversity and improves interference management and sensing coverage.
	For example,~\cite{re:lae-multipoint-2025-6} jointly optimizes coordinated transmit beamforming, authorized-UAV trajectories, and base station--UAV associations to maximize the average sum rate of authorized UAVs while guaranteeing the minimum illumination power over the target 3D sensing region.
	In~\cite{re:lae-isac-resource-2026-4}, transmit beamforming, receive filtering, and UAV trajectories are jointly designed to maximize the weighted sum rate under sensing-SINR constraints.
	However, networked cooperation also introduces inter-base-station interference and tighter coupling among beamforming, association, receive filtering, and trajectory variables, leading to mixed-integer or highly non-convex optimization problems.
	Therefore, the performance improvement offered by base station cooperation is accompanied by increased design and computational complexity.

	\item 
	\textbf{Aircraft acting as aerial mobile base stations:}
	Some works focus on UAVs acting as aerial mobile base stations, where UAVs communicate with terrestrial users and achieve the sensing of terrestrial targets.
	In~\cite{re:lae-isac-resource-2023-5}, an integrated periodic sensing and communication mechanism is proposed, where the UAV trajectory, user association, target-sensing selection, and transmit beamforming are jointly optimized to maximize the achievable communication rate under sensing-frequency and beampattern-gain requirements.
	Instead of requiring sensing in every communication slot, this mechanism periodically selects sensing times according to the sensing-frequency requirement, thereby avoiding excessive sensing and the resulting waste of spectrum and energy resources.
	In~\cite{re:lae-isac-resource-2023-2}, an adaptable ISAC mechanism is further developed to perform sensing on demand and flexibly configure the sensing duration according to application requirements.
	Different from the above studies that mainly treat sensing performance as a constraint,~\cite{re:lae-isac-resource-2024-1} formulates a normalized weighted objective to jointly optimize the UAV trajectory and user bandwidth allocation under an onboard-energy constraint.
	By varying the weighting factor, different communication--sensing operating points can be obtained, thereby presenting their performance tradeoff more directly.
	Moreover, in multi-UAV-assisted ISAC systems, trajectory design becomes more tightly coupled with user association and radio-resource allocation.
	In~\cite{re:lae-isac-resource-2024-6}, multiple UAV trajectories, user association, and beamforming are jointly optimized to maximize the sum weighted bit rate of all ground users while guaranteeing the sensing beampattern gain of the target.
	In contrast,~\cite{re:lae-isac-resource-2024-7} introduces radar mutual information (MI) as the sensing metric and jointly optimizes node scheduling, transmit power, and 3D UAV trajectories to maximize the minimum communication rate per IoT node under radar-MI constraints.
\end{itemize}

Existing studies have demonstrated that aircraft trajectories can be treated as controllable spatial resources and jointly optimized with communication and sensing variables.
However, most of the reviewed formulations consider a finite mission period and rely on deterministic LoS channel models and known or estimated user and target states.
As the numbers of users, targets, and cooperative base stations increase, model-based methods using alternating optimization, semidefinite relaxation, and successive convex approximation face rapidly increasing computational complexity, whereas learning-based methods require extensive interaction data and training overhead.
Moreover, beamforming, power allocation, and scheduling can be updated on a much shorter time scale than a physical aircraft trajectory.
How to achieve scalable online trajectory and resource optimization under imperfect channel information, uncertain target motion, limited onboard energy, residual interference, and clutter therefore remains an important problem for practical low-altitude ISAC.

\subsection{Summary}
Existing low-altitude ISAC studies have established a substantial technical foundation covering system architectures, channel and signal models, waveform and receiver design, resource allocation, and aircraft trajectory optimization. 
Nevertheless, ISAC lacks identity awareness: estimates of an aircraft's position, velocity, and trajectory cannot directly determine who it is. 
The next section therefore discusses ISCAI by incorporating identification into ISAC.

\section{ISCAI for LAINs}
\label{sec:low-ISCAI}
In this section, we introduce ISCAI for LAINs by integrating communication, sensing, and identification into a unified framework.
Low-altitude identification broadly includes target classification and individual identity acquisition, and this section primarily focuses on the latter.
First, we discuss the evolution from ISAC to ISCAI, and then propose the BackCom-based identification paradigm, serving as a supplement to existing identification systems.
Furthermore, we integrate it into existing low-altitude sensing and ISAC systems to form ISAI and ISCAI systems, respectively, and explore their system models, signal models, multifunctional receiver design, and mutual integration gains in depth.

\subsection{From ISAC to ISCAI}
In this paper, ISCAI refers to an integrated system where communication, sensing, and identification can share the same waveform, spectrum, and hardware resources, and are jointly designed in terms of waveform, resource allocation, and receiver processing so as to obtain mutual integration gains.
The defining requirement is that identification-related information or variables are explicitly considered together with communication symbols and sensing parameters in system modeling, signal processing, receiver design, and system optimization.
Further, the evolution from ISAC to ISCAI can be explained from two aspects.

\begin{itemize}
	\item
	\textbf{Inherent functional limitation of ISAC:}
	Low-altitude ISAC possesses an inherent functional limitation, where it can only estimate physical state information like the positions, velocities, and trajectories of targets without the ability to acquire their identity information.
	However, with the increasing need for airspace safety and regulatory compliance, acquiring targets' identity information is becoming increasingly essential.
	Taking airspace surveillance as an example, LAINs need to promptly acquire the identities of various targets to support differentiated control measures such as prioritized tracking, early warnings, or countermeasures.
	
	\item
	\textbf{Benefits of ISCAI for LAINs:}
	ISCAI can be viewed as the further evolution of ISAC, and enables LAINs to simultaneously possess target sensing and identification capabilities by deeply integrating identification into ISAC.
	The acquired identity information can support subsequent authentication, regulatory authorization, trajectory association, and differentiated airspace management.
	In short, low-altitude ISAC addresses the problem of making targets visible, while ISCAI further aims to achieve the goals of seeing clearly, distinguishing accurately, and controlling effectively, representing an important evolutionary direction for LAINs.
\end{itemize}

Furthermore, we observe that the broadcast-based identity reporting, network-based identity reporting, and RFF-based identification paradigms discussed in Section 2.3 all rely on signals actively transmitted by aircraft.
Broadcast-based and network-based identity reporting carry identity identifiers as data payloads over aircraft-initiated broadcast or uplink communication links, whereas RFF-based identification infers device identities from physical-layer hardware characteristics embedded in signals transmitted by aircraft.
In contrast, typical monostatic and distributed ISAC systems generally operate by having the infrastructure transmit communication or sensing signals, while targets merely reflect the incident signals.
Because of the differences in signal origins, propagation links, and echo-processing procedures, existing identification paradigms can be deployed in parallel with ISAC, but they cannot naturally share waveforms, spectrum, and hardware resources with ISAC systems.
The simple parallel deployment of ISAC and identity reporting, in which the aircraft identifier is treated only as an ordinary communication payload and is not coupled with the ISAC system model or signal processing, is therefore not regarded as ISCAI.

To address this integration problem, the following subsection introduces BackCom-based identification as a complement to existing identification paradigms and regards it as the key technical bridge from existing ISAC toward ISAI and ISCAI.
Accordingly, this paper focuses on BackCom-based realizations of ISAI and ISCAI.
It should be noted that the definition of ISCAI is not restricted to any particular identification mechanism, and this paper does not exclude the possibility of other ISCAI realizations, which are left for future exploration.

\subsection{BackCom-based identification paradigm}
Inspired by the paradigm where ground traffic control systems capture vehicle license plates through visual sensors to subsequently complete vehicle identification, this section introduces BDs as electronic license plates for aircraft to propose a novel BackCom-based aircraft identification paradigm.

\subsubsection{The basic principles of BackCom}

\begin{figure*}[!t]
	\centering
	\includegraphics[width=0.9\textwidth]{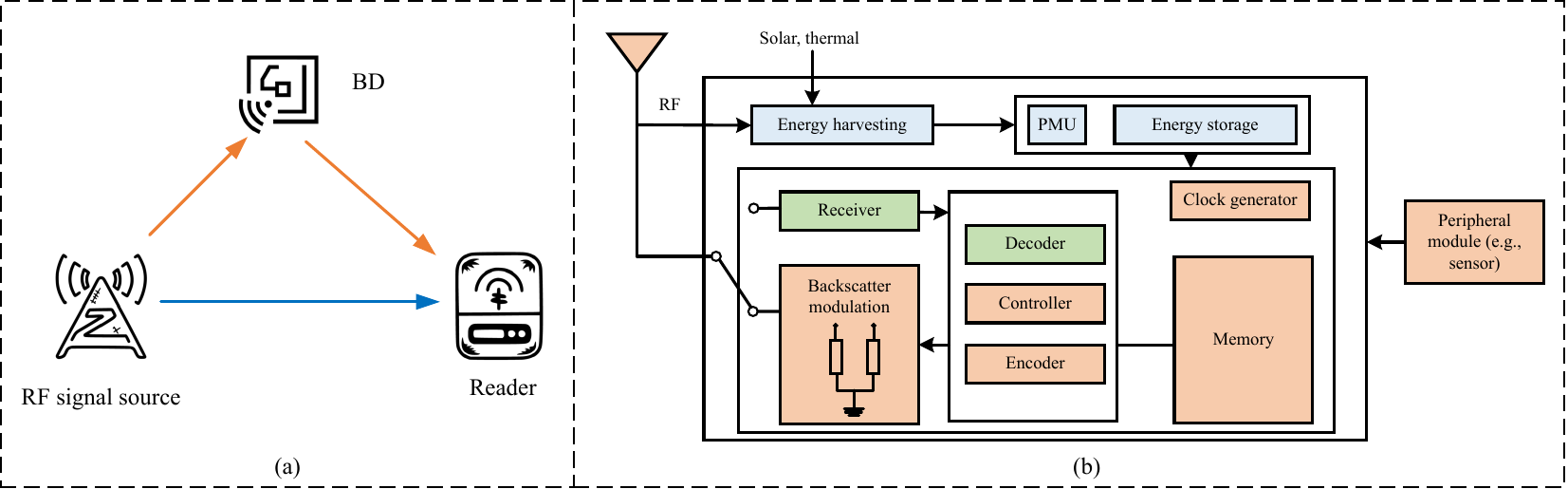}
	\caption{BackCom system: (a) Bistatic BackCom configuration comprising an RF signal source, a BD, and a reader; (b) Internal structure of the BD, mainly consisting of the backscatter modulation, controller, memory, encoder, energy harvesting, power management unit (PMU), and energy storage modules.}
	\label{fig:backcom-bd}
\end{figure*}

BackCom is a passive or semi-passive communication paradigm in which the BD transmits information by modulating and reflecting an incident RF signal, rather than generating its own carrier wave~\cite{re:lae-BC-2022-8,re:lae-BC-2024-7,re:lae-BC-2026-16}. 
As shown in Figure~\ref{fig:backcom-bd}(a), a typical BackCom system comprises three parts, namely an RF signal source, a BD, and a reader.
As the core device of BackCom, the BD primarily contains the backscatter modulation, controller, memory, encoder, energy harvesting, power management unit, and energy storage modules, as shown in Figure~\ref{fig:backcom-bd}(b).
The BackCom process proceeds as follows.
\begin{itemize}
	\item
	\textbf{RF source signal:}
	By considering a continuous-wave or information-bearing carrier transmitted by an RF source, the RF source signal $s_{\mathrm{rf}}(t)$ is expressed as
	\begin{equation}
		s_{\mathrm{rf}}(t) = \sqrt{P_t}\, x(t)\, e^{j2\pi f_c t},
	\end{equation}
	where $P_t$ denotes the transmit power, $x(t)$ is the baseband waveform, and $f_c$ is the carrier frequency.
	
	\item
	\textbf{Backscatter modulation:}
	Backscatter modulation aims to establish mapping rules between transmitted symbols and reflection coefficients~\cite{re:lae-BC-2018-1}.  
	The BD modulates information onto the reflected signal by switching its antenna load impedance among a finite set of states $\{Z_1, Z_2, \ldots, Z_I\}$. 
	Each impedance state produces a distinct reflection coefficient
	\begin{equation}
		\Gamma_i = \frac{Z_i - Z_A^*}{Z_i + Z_A},
		\quad i = 1, 2, \ldots, I,
	\end{equation}
	where $Z_A$ is the antenna impedance of the BD. 
	By switching among these states according to a pre-stored information symbol sequence $\{\omega_n\}$, the BD produces a time-varying reflection coefficient $\Gamma(t)$, which acts as a passive amplitude-and-phase modulator. 
	
	\item
	\textbf{Backscattered signal:}
	The backscattered signal received at the reader can be written as 
	\begin{equation}
		r(t) = h_{f}\, h_{g}\, \Gamma(t)\, s_{\mathrm{rf}}(t) + z(t),
	\end{equation}
	where $h_{f}$ denotes the channel coefficient from the RF source to the BD, $h_{g} $ denotes the channel coefficient from the BD to the reader, and $z(t)$ is the AWGN. 
	Then, the reader performs down-conversion and analog-to-digital conversion on $r(t)$, and utilizes classical detectors~\cite{re:lae-BC-2016-10,re:lae-BC-2017-11} such as energy detectors, coherent detectors, or differential detectors to detect the transmitted BD symbol sequence $\{\omega_n\}$.
\end{itemize}

In this process, the BD does not generate the RF carrier itself, which eliminates power-intensive components such as digital-to-analog converters and power amplifiers, and the BD can directly harvest energy from ambient electromagnetic waves through the energy harvesting module.
This mode dramatically reduces the energy consumption of BackCom, making it well suited for energy-constrained IoT applications~\cite{re:lae-BC-2023-2}, such as smart logistics and intelligent warehousing.

\subsubsection{The process of BackCom-based identification}

Directly installing physical license plates on small or medium aircraft increases wind resistance and body weight, thereby affecting their aerodynamic structures, which consequently severely constrains the endurance and maneuvering performance of aircraft.
In contrast, BDs can be attached to aircraft surfaces, forming aircraft-BD pairs.
BDs can serve as lightweight electronic license plates, where their internally pre-stored information sequences contain key identity information like the unique identifiers and registered locations of aircraft~\cite{re:lae-isai-2026-1}.
\begin{itemize}
	\item
	\textbf{Lightweight and low-power device:}
	Featuring light weight and extremely low power consumption, BDs can be directly attached to aircraft surfaces without imposing a significant payload burden, and passive BDs are able to autonomously maintain operations by harvesting RF energy without requiring external power supplies.
	
	\item
	\textbf{Controlled identity information transmission:}
	BackCom supports controlled communication, in which BDs can be configured to initiate information transmission only after being activated by external wake-up signals~\cite{re:lae-BC-2021-9}.
	This externally triggered mechanism aligns well with the requirements of airspace surveillance.
\end{itemize}

The BackCom-based identification process is described as follows.
When LAINs need to execute identification on aircraft within the airspace, the identification station (acting as the RF source) first transmits wake-up signals to activate BDs on aircraft surfaces, and then transmits excitation carriers.
Then, BDs modulate and reflect incident signals according to pre-stored information sequences, while the receiver demodulates the symbol sequences transmitted by the BDs from the reflected signals, extracts aircraft identity information, and finally compares it with local databases to complete aircraft identification. 
BDs can also interconnect with aircraft sensors to transmit additional data such as aircraft state parameters along with the surrounding temperature and humidity~\cite{re:lae-BC-2023-2}, further enriching the information dimensions of airspace surveillance.
Although this identification paradigm shares fundamental similarities with conventional radio frequency identification (RFID) technology, their primary difference is that an RFID system typically operates as an independent system requiring dedicated excitation sources.
The BackCom-based identification paradigm can directly reuse the RF signals of existing communication systems~\cite{re:lae-BC-2013-13} or sensing systems~\cite{re:fmcw-bd-2015}, thereby effectively reducing spectrum overhead and hardware deployment costs.

It should be clarified that BackCom-based identification primarily performs identity-information acquisition, whereas the authenticity of the acquired identity and the operational legitimacy of the aircraft still need to be determined through subsequent authentication and regulatory authorization mechanisms.
For example, for a target equipped with a forged or cloned BD, successfully demodulating the BD sequence only means that the system has acquired the identity claimed by the target.
It does not, by itself, prove that the identity is authentic, has not been cloned, or is authorized to operate in the current airspace.
Nevertheless, compared with standalone identity reporting, the joint processing of ISAI and ISCAI in Sections 4.3 and 4.4 provides additional consistency constraints for detecting identity spoofing or cloning.
Because the BD identity symbols and the target parameters, including delay, Doppler shift, and angles, are jointly recovered from the same echo, the acquired identity is naturally associated with the physical location and trajectory of the target.
If the same identity is simultaneously detected at different locations, or is associated with clearly inconsistent observed trajectories, particularly under distributed multi-station observations, the inconsistency between the identity and the physical states can serve as an important indicator of identity cloning or spoofing.

\subsubsection{Identification performance metrics}

Since identification is treated as a design dimension in its own right, it also requires its own performance metrics, in parallel with the communication and sensing metrics summarized in Section 3.1.4.
Identification performance can generally be characterized from two aspects, namely decision reliability and timeliness.
In terms of decision reliability, the missed-identification probability describes the probability that the system fails to produce a valid identity decision for a target whose identity is to be acquired; the identity false alarm probability describes the probability that a valid identity is declared for a target that provides none; and the false-identification probability describes the probability that the identity reported by the system does not match the target's true identity.
Timeliness is described by the identification latency, i.e., the time from initiating identification to producing an identity decision, which is a key constraint in airspace-regulation scenarios.
For the BackCom-based realization considered in this paper, these metrics take a concrete form.
Missed identification arises mainly from two-way link loss, blockage, fading, and BD synchronization errors, all of which prevent the BD symbol sequence from being detected.
An identity false alarm occurs when residual clutter or noise in the echo of a BD-free target is demodulated into a valid codeword.
The corresponding identification latency consists mainly of the BD wake-up and synchronization time, the identity-sequence transmission time, and the joint parameter-estimation and symbol-detection time.

\subsubsection{The main advantages and limitations of BackCom-based identification}

This subsection further discusses the main advantages and limitations of this paradigm.
\begin{itemize}
	\item
	\textbf{Main advantages:}
	First, BackCom-based identification can reuse existing RF signals to transmit aircraft identity information.
	Unlike broadcast-based and network-based identity-reporting mechanisms, this paradigm does not require a dedicated identification waveform or additional spectrum resources.
	Second, as discussed in Section 4.1, unlike broadcast-based and network-based active identity-reporting mechanisms and RFF-based identification, the passive reflection architecture of BackCom is highly compatible with the signal transmission and reception procedures of typical ISAC systems.
	When BackCom-based identification is incorporated into such a system, the ISAC transmitter can simultaneously serve as the RF excitation source for the BD.
	The aircraft-mounted BD embeds its identity sequence into the target echo through backscatter modulation, while the ISAC receiver recovers the BD identity symbols and estimates the target parameters, including delay, Doppler shift, and angles, from the same received signal.
	Therefore, BackCom-based identification does not need to operate as an independent system in parallel with ISAC, but can instead be directly incorporated into the ISAC framework.
	This compatibility at both the signal and system-architecture levels makes BackCom-based identification a key identification paradigm for realizing ISAI and ISCAI.
	Finally, BackCom-based identification endows LAINs with the capability to actively acquire aircraft identities on demand, overcoming the limitation of existing identification systems that can only passively wait for aircraft reporting.
	Broadcast-based and network-based identity-reporting mechanisms require the system to wait for aircraft to actively upload their identity information, while RFF-based identification can extract device-specific features only when the target is actively transmitting RF signals.
	
	\item
	\textbf{Main limitations:}
	First, BackCom-based identification requires aircraft to be equipped with BDs in advance.
	In contrast, network-based identity reporting can use the existing active communication equipment of aircraft, while RFF-based identification can directly analyze the RF signals already transmitted by targets without requiring the installation of additional BDs.
	Second, the backscattered signal must propagate over both the forward link from the infrastructure to the BD and the backward link from the BD to the receiver, thereby suffering from substantial round-trip propagation loss.
	Compared with active-transmission-based identity-reporting mechanisms such as ADS-B and Remote ID, BackCom-based identification generally has a shorter effective identification range.
	Under blockage or long-distance conditions, the received BD signal may be too weak to support reliable BD-symbol detection, thereby increasing the probability of missed identification.
	Third, the BD must first be awakened and synchronized with the incident signal, which may introduce additional wake-up and synchronization latency compared with RFF-based identification. 
	Moreover, compared with broadcast-based and network-based active identity-reporting mechanisms, the passive reflection link supports only a limited data transmission rate.	
\end{itemize}

\subsection{Integrated sensing and identification}
Building on the BackCom-based identification paradigm in Section 4.2, this section integrates it into low-altitude sensing to form the ISAI paradigm, aiming to simultaneously achieve target sensing and individual identification while sharing spectrum and hardware resources.
ISAI can be viewed as a functional subset of ISCAI, where ISCAI naturally degenerates into ISAI when LAINs turn off the communication function.
It is well suited for scenarios such as passive surveillance where silent targets do not communicate with surveillance stations.

\subsubsection{System model}

\begin{figure*}[!t]
	\centering
	\includegraphics[width=0.8\textwidth]{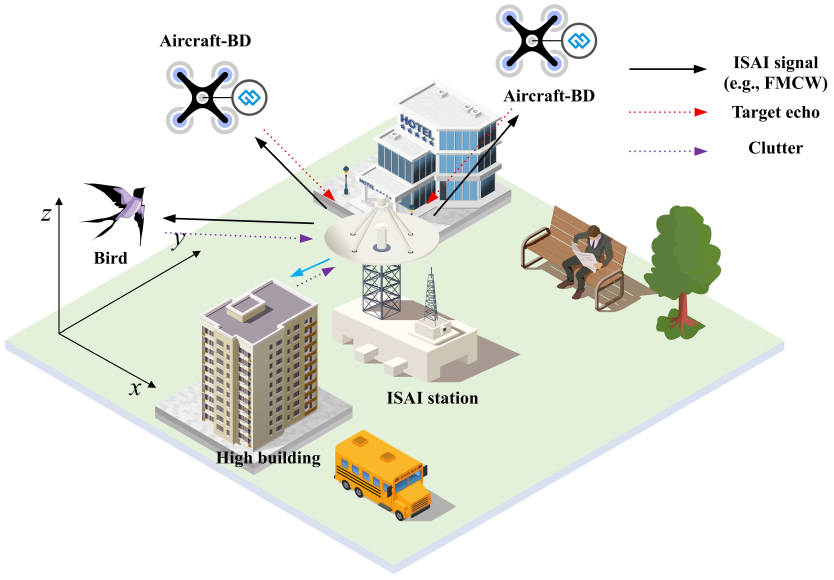}
	\caption{ISAI system model, where BDs are attached to aircraft as electronic license plates.
		An ISAI station emits sensing signals to illuminate aircraft-BD pairs, and receives the reflected signals, from which the ISAI station simultaneously estimates aircraft physical parameters and extracts their identity information.}
	\label{fig:isai-system}
\end{figure*}

As shown in Figure~\ref{fig:isai-system}, aircraft fly within the coverage of an ISAI station.
Driven by regulatory demands, the ISAI station localizes aircraft and measures their velocities while acquiring their identity information and associating it with their physical trajectories, thereby providing inputs for subsequent authentication and regulatory authorization.
The ISAI station first transmits wake-up signals to activate the BDs, and then emits sensing signals such as FMCW signals~\cite{re:lae-radar-2005-1}, where sensing signals are simultaneously utilized for target identification.
Then, BDs embed aircraft identity information into echo signals through backscatter modulation. 
Upon receiving echoes along with environmental clutter (e.g., the reflected signals from birds), the ISAI station performs clutter suppression, and then conducts joint parameter estimation and BD symbol detection to estimate the physical parameters of aircraft, while simultaneously demodulating BD symbols to extract identity information~\cite{re:lae-isai-2026-1,re:fmcw-bd-survey-2025}.
Similar to ISAC, ISAI can also be categorized into three architectures: monostatic, bistatic, and distributed.

It is worth noting that, for targets without BDs, the ISAI station can also receive their reflected echoes, and can still perform detection, localization, velocity estimation, and tracking based on these structural reflection echoes, although their individual identities cannot be acquired through BackCom.
Since the echoes from such targets consist solely of surface reflection components, their equivalent backscattering coefficients remain approximately constant throughout the observation period, inherently lacking the symbol-level time-varying characteristics introduced by BD modulation. 
Under the ideal signal model considered in this subsection, executing BD symbol detection on such an echo yields a constant sequence that does not match any entry in the identity database.
Besides, in practice, a BD-equipped target may still fail to produce a detectable sequence under two-way propagation loss, blockage, fading, or synchronization errors.
Therefore, the absence of a valid BD sequence only indicates that the system has failed to acquire the target identity during the current observation.
When no valid identity has been acquired during the current observation, the ISAI station should trigger complementary identification methods and continued observation, such as repeated interrogation, continuous tracking, and early warning.

\subsubsection{Signal model}
\label{sec:4-3-2}
In this subsection, we take the example of the FMCW-enabled monostatic ISAI system to elaborate the signal models of the whole process mentioned above, where the ISAI station utilizes a transmitter equipped with a single antenna to emit the FMCW signal, while employing a receiver equipped with a UPA to receive echoes from $Q$ aircraft-BDs.
Here, to emphasize the effect of introducing BDs on system design, we assume that static and dynamic clutter, alongside self-interference, have been effectively suppressed. 
Specifically, since the introduction of BDs does not alter the clutter channel in~\eqref{eq:3-1-2-3}, clutter suppression can follow the schemes reviewed in Section 3.3.3, while self-interference cancellation can leverage existing full-duplex system techniques~\cite{re:lae-isac-full-2014}.
The specific signal models are described as follows.

\begin{itemize}
	\item
	\textbf{Emitted FMCW signal:}
	Following the definitions of FMCW system parameters in Section 3.3.1, the ISAI station emits $K$ FMCW chirps, each with a duration of $T_c$, where the specific signal expression is given by $x_b(t)$ in \eqref{eq:3-3-1-1}.
	
	\item
	\textbf{Transmitted waveforms of BDs:}
	BDs support diverse modulation schemes, such as amplitude shift keying, PSK, and FSK. Herein, PSK is selected as a representative example to elaborate on the transmitted waveforms of BDs.
	Let $b_q(t)$ represent the waveform of the $q$-th BD, which comprises $L$ BD symbols within each FMCW period $T_c$. 
	Accordingly, the duration of each BD symbol, denoted by $T_b$, is given by $T_b = T_c / L$. 
	Then, the transmitted waveform $b_q(t)$ can be written as
	\begin{align}
		b_q(t)= \sum_{k=0}^{K-1} \sum_{l=0}^{L-1} \omega_{q,k,l}
		\textrm{rect} \left( \frac{t-kT_{c}-lT_{b}}{T_b}\right),
		\label{eq:fmcw-bd}
	\end{align}
	where $\omega_{q,k,l}$ denotes the $l$-th symbol transmitted by the $q$-th BD during the $k$-th FMCW period.
	
	\item
	\textbf{Echo signal model:}
	Echoes comprise two components: the backscattered component originating from the BD antennas and the reflected component resulting from the physical surfaces of aircraft-BDs~\cite{re:lae-BC-2020-15}.
	Then, they can be further expressed as
	\begin{align}
		\mathbf{r}(t)=\sum_{q=0}^{Q-1}
		x_b\left( t-\tau_q\right)  h_q^{\mathrm{sen}}\left( \beta_q^{\mathrm{bd}}b_q\left( t-\tau_q\right) +\beta_q^{\mathrm{str}}\right)e^{j2\pi \nu_q t }
		\mathbf{a}_r (\theta_q,\phi_q)+ \mathbf{z}_1^{\mathrm{sen}}(t),
		\label{eq:fmcw-echo}
	\end{align}
	where $\theta_q$, $\phi_q$, $\tau_q$, and $\nu_q$ denote the azimuth angle, elevation angle, two-way delay, and two-way Doppler frequency shift associated with the $q$-th aircraft-BD, respectively; $h_q^{\mathrm{sen}}$ is the two-way propagation gain; $\beta_q^{\mathrm{bd}}$ and $\beta_q^{\mathrm{str}}$ denote the backscatter coefficient associated with the BD antenna and the structural reflection coefficient associated with the physical surface of the $q$-th aircraft-BD, respectively; $\mathbf{z}_1^{\mathrm{sen}}(t)$ represents the equivalent noise including the AWGN and the residual interference following self-interference and clutter suppression.
	
	\item
	\textbf{Discrete beat signal model:}
	The beat signal $\mathbf{y}(t)$ is then obtained by the mixing operation, and is uniformly sampled with an interval of $T_s$ to obtain the discrete beat signal $\mathbf{y}_{k,l,n}$.
	According to \eqref{eq:3-3-3-1} in Section 3.3.3, $\mathbf{y}_{k,l,n}$ can be written as
	\begin{align}
		\mathbf{y}_{k,l,n}
		=\sum_{q=0}^{Q-1} h_q^{\mathrm{sen}}
		e^{-j2\pi f_s\tau_q}
		e^{-j2\pi\left( \mu \tau_q-\nu_q\right) (lT_b+nT_s)} 
		e^{j2\pi \nu_q k T_c}
		\zeta_{q,k,l,n}
		\mathbf{a}_r(\theta_q,\phi_q)+\mathbf{z}_{k,l,n}^{\mathrm{sen}},
		\label{eq:fmcw-beat-discrete}
	\end{align}
	where $\zeta_{q,k,l,n}=\left( \beta_q^{\mathrm{bd}} b_q\left( t-\tau_q\right)+ \beta_q^{\mathrm{str}}\right)|_{t = kT_c + lT_b + nT_s}$; $\mathbf{z}_{k,l,n}^{\mathrm{sen}}=\mathbf{z}^{\mathrm{sen}}(t)|_{t = kT_c + lT_b + nT_s}$; $\mathbf{z}^{\mathrm{sen}}(t)$ denotes the equivalent noise component after the mixing operation.
\end{itemize}

Comparing~\eqref{eq:fmcw-beat-discrete} with the conventional FMCW beat signal in~\eqref{eq:3-3-3-1}, the composite reflection term $\zeta_{q,k,l,n}$ is the structural difference introduced by the identification function.
When the BD stops transmitting its identity sequence and holds its reflection state fixed, $b_q(t)$ in~\eqref{eq:fmcw-bd} becomes constant, so that $\zeta_{q,k,l,n}$ degenerates into a time-invariant reflection coefficient and~\eqref{eq:fmcw-beat-discrete} recovers~\eqref{eq:3-3-3-1}.
In addition, it should be noted that the above signal model treats the residual clutter and residual self-interference as part of the equivalent noise, both of which degrade the identification performance.
The residual self-interference reduces the effective SINR of BD symbol detection, thereby increasing the false-identification probability, while time-varying residual clutter may also cause identity false alarms.
Moreover, since the residual interference scales with the transmit power, this degradation cannot be eliminated simply by increasing the transmit power.

\subsubsection{Joint sensing and identification receiver design}

Based on the discrete beat signal model $\mathbf{y}_{k,l,n}$ in \eqref{eq:fmcw-beat-discrete}, the ISAI station needs to design a joint parameter estimation and symbol detection framework to estimate the azimuth angles, elevation angles, and delays/ranges of aircraft for 3D localization, while simultaneously estimating their Doppler frequencies/radial velocities and detecting the BD symbols to acquire aircraft identity information.

Compared with the received signal model of FMCW-enabled sensing in \eqref{eq:3-3-3-1} of Section~3.3.3, the integration of the identification function introduces additional BD symbol terms $\{\zeta_{q,k,l,n}\}$ into $\mathbf{y}_{k,l,n}$, fundamentally altering the signal structure and posing the following core challenges for joint sensing and identification receiver design.

\begin{itemize}
	\item
	\textbf{The first challenge is the bilinear coupling between BD symbol terms and aircraft parameters.}
	As can be seen from \eqref{eq:fmcw-beat-discrete}, 
	$\zeta_{q,k,l,n}$ consists of a product term $\beta_q^{\mathrm{bd}}\omega_{q,k,l}$ coupling the BD symbol with the backscatter coefficient, plus a structural reflection term $\beta_q^{\mathrm{str}}$, causing the BD symbols to be tightly coupled with the aircraft's delay and Doppler frequency parameters in the fast-time dimension, as well as with the aircraft's Doppler frequency in the slow-time dimension.
	This coupling exhibits a bilinear structure: given aircraft parameters, BD symbol detection reduces to a linear problem; conversely, given BD symbols, aircraft parameter estimation simplifies to a standard spectral estimation problem.
	However, when both are simultaneously unknown, the problem exhibits an inherently nonlinear nature, rendering widely used methods mentioned in Section 3.3.3, such as cascaded FFT processing, no longer directly applicable.
	
	\item
	\textbf{The second challenge is the synchronization errors between the BD and the incident sensing signal, as well as the limited clock precision of the BD itself.}
	On the one hand, the BD needs to synchronize with the incident sensing signal to ensure that symbol boundaries are aligned with chirp periods, while the low-power design of the BD limits the precision of this synchronization process, inevitably introducing synchronization errors.
	On the other hand, the internal clock generator of the BD itself also suffers from insufficient precision, causing frequency offsets and timing drifts during continuous operation.
	These errors jointly degrade the accuracy of parameter estimation and symbol detection, posing additional challenges for joint sensing and identification receiver design.
\end{itemize}

To address the above problems,~\cite{re:fmcw-bd-2015} applies a line code in which the BD transmits a symbol block during one chirp and its negated copy during the next chirp, so that static clutter, which is identical across chirps, is removed by chirp-pair subtraction, and further derives an upper bound on the achievable communication rate for a given node density and radar configuration.
\cite{re:fmcw-bd-2018} adopts an architecture with a single transmit antenna and multiple receive antennas, and jointly estimates the delays and AoAs of multiple BDs for localization based on the coherent reflected responses of switched injection-locked oscillators.
\cite{re:fmcw-bd-2024} encodes downlink information at the radar through chirp-slope-shift keying, while the BD conveys uplink information through backscatter modulation, thereby supporting ranging and bidirectional BackCom over the same FMCW link.
The above works mainly consider static or slowly moving BDs, where the BD modulation and the target range can be processed separately, e.g., the BD modulation spans multiple chirps and is extracted along the slow-time dimension in~\cite{re:fmcw-bd-2024}, whereas the BD data are demodulated around the estimated range-dependent beat frequency in~\cite{re:fmcw-bd-2015}.
Moreover, the above works model only the antenna-mode backscatter of the BD, while neglecting the structural reflection component $\beta_q^{\mathrm{str}}$ that does not vary with the symbols.

\cite{re:isabc-ris-fmcw-2025},~\cite{re:lae-isai-2026-1}, and~\cite{re:lae-isai-2026-4} further investigate the joint recovery problem in moving scenarios.
\cite{re:isabc-ris-fmcw-2025} considers an FMCW system in which an RIS serves as the reflective modulator, and proposes an element-wise complex division algorithm that directly decouples the sensing parameters from the reflection-modulated symbols, where the estimated sensing parameters are further exploited to assist symbol demodulation and thus reduce the pilot overhead.
In~\cite{re:lae-isai-2026-1}, a zero-padded BD symbol pattern is proposed to mitigate inter-symbol interference caused by transmission delays and synchronization errors.
Further, by employing the proposed BD symbol pattern and a truncation operation, the discrete beat signals are reconstructed into a low-rank matrix model, and then a matrix decomposition-based two-stage scheme is proposed to address the tight coupling between aircraft parameters and BD symbols.
In contrast, a tensor-based multi-antenna framework is proposed in~\cite{re:lae-isai-2026-4}.
In this work, BDs adopt a differential encoding strategy between adjacent BD symbols in each FMCW period.
Then, at the receiver, $\mathbf{y}_{k,l,n}$ is first reshaped into a fourth-order tensor, followed by the design of a non-iterative tensor decomposition algorithm based on singular value decomposition and eigenvalue decomposition.
Then, by utilizing amplitude characteristics of the digital sinc function and an energy detection-based differential decoding scheme, a computationally efficient scheme is proposed to achieve 3D localization of aircraft, estimate their radial velocities, and detect BD symbols from the four estimated factor matrices.

\subsubsection{Mutual integration gains}
\label{sec:4-3-3}

In ISAI, sensing and identification achieve complete spectrum and hardware sharing, while targets' physical parameters and identity information can be automatically associated. 
Their deep integration brings the following significant mutual integration gains. 
\begin{itemize}
	\item
	The identification function can directly reuse the existing sensing signal as the RF carrier for BackCom to complete aircraft identification without additional resource overhead. 
	This seamless integration eliminates the need for dedicated identification waveforms or separate frequency allocations, making the proposed scheme particularly attractive for spectrum-constrained and cost-sensitive deployments.
	
	\item
	BD antennas specifically designed for the sensing frequency band typically have stronger reflection capabilities than the aircraft's own materials, effectively enlarging the equivalent RCS of aircraft.
	As depicted in~\cite{re:lae-isai-2026-4}, the CRLBs for parameter estimation can be effectively decreased after the introduction of BDs, thereby improving the achievable sensing accuracy.
	This enhancement is especially beneficial for small or low-observable aircraft, where the natural RCS may be insufficient for reliable detection at extended ranges.
	
	\item
	Identification results can directly guide sensing resource allocation in an adaptive and priority-driven manner. Once a target is classified as high-threat, the system can automatically increase its revisit frequency, assign higher-resolution sensors, or dedicate additional tracking channels to maintain continuous situational awareness. Conversely, targets identified as low-priority can be monitored with reduced resources, freeing up capacity for more critical observations.
	
	\item
	Identity information can facilitate data association for multiple targets in distributed architectures, as reliably decoded identifiers provide additional cues for associating observations across stations. However, missed BD detections or decoding errors may still lead to association failures or incorrect associations, so position and trajectory information should also be considered.
	
\end{itemize}

\subsection{Integrated sensing, communications, and identification}

Building on the BackCom-based identification paradigm, this section integrates it into low-altitude ISAC to form the ISCAI paradigm, aiming to simultaneously achieve communication data transmission, target sensing, and individual identity acquisition while sharing spectrum and hardware resources.

\subsubsection{System model}

\begin{figure*}[!t]
	\centering
	\includegraphics[width=0.8\textwidth]{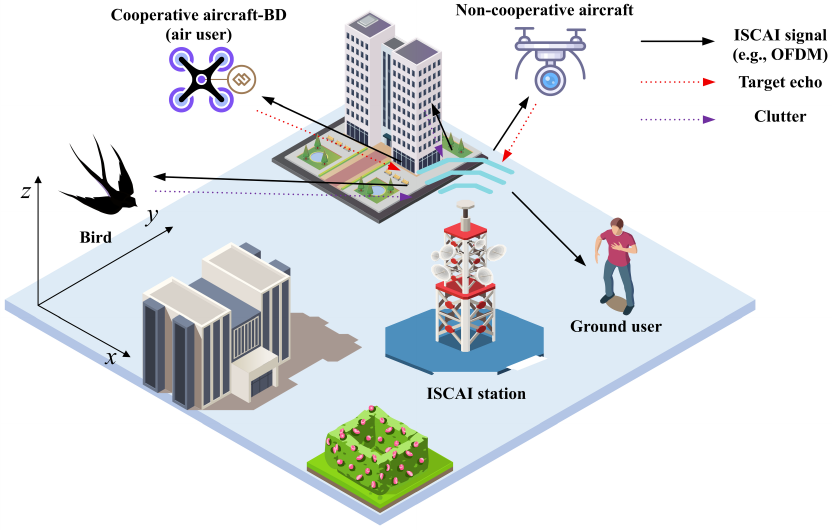}
	\caption{ISCAI system model, where BDs are attached to aircraft as electronic license plates.
	The ISCAI station transmits communication signals to serve cooperative aircraft-BDs and ground users, and receives echoes, from which the ISCAI station completes aircraft physical parameter estimation and identity information extraction.}
	\label{fig:ISCAI-system}
\end{figure*}

As shown in Figure~\ref{fig:ISCAI-system}, cooperative and non-cooperative aircraft fly simultaneously within the coverage of an ISCAI station.
The ISCAI station communicates with cooperative aircraft and ground users for information exchange, while localizing aircraft and measuring their velocities to track all aircraft in the airspace, acquiring their identity information, and associating the information with their physical trajectories.
The specific operational procedures of the ISCAI system are detailed as follows.
\begin{itemize}
	\item
	The ISCAI station first emits wake-up signals to activate the BDs, and then transmits communication signals, such as OFDM and OTFS signals.
	Communication signals are simultaneously utilized for aircraft sensing and identification to act as triple-function signals.
	
	\item Upon receiving the downlink communication signals, cooperative aircraft and terrestrial users execute advanced channel estimation and signal detection algorithms to mitigate multipath fading and the severe Doppler shifts induced by high mobility, thereby reliably recovering the transmitted data streams.
	
	\item
	The communication signal illuminates all aircraft-BD pairs in the low-altitude region, generating two components in echoes: a component directly reflected by the physical surfaces of aircraft-BDs, as well as a component carrying identity information and produced by BDs switching their antenna load impedances according to their identity sequences.
	
	\item
	Upon capturing echoes and environmental clutter, the ISCAI station simultaneously estimates the physical parameters of aircraft including positions and velocities for sensing, and detects BD symbol sequences to extract and match identity information for identification.
\end{itemize}

\subsubsection{Signal model}
\label{sec:4-4-2}
In this subsection, we take the example of the OFDM-enabled monostatic ISCAI system to elaborate the signal models of the whole process mentioned above~\cite{re:lae-iscai-2023-2,re:isabc-shao-2024}.
The ISCAI station employs UPAs to transmit the downlink OFDM signal to serve $P$ single-antenna communication users, and to receive echo signals for joint sensing and identification of $Q$ aircraft-BDs.
Similar to Section \ref{sec:4-3-2}, clutter and self-interference are assumed to be effectively suppressed via the clutter suppression schemes reviewed in Section 3.2.2 and full-duplex system techniques~\cite{re:lae-isac-full-2014}, respectively.
The specific signal models are described as follows.

\begin{itemize}
	\item
	\textbf{Transmitted OFDM signal:}
	Following the definitions of OFDM system parameters in Section 3.2.1, the ISCAI station transmits a frame of $N$ OFDM symbols, where the specific signal expression is given by $\mathbf{s}(t)$ in \eqref{eq:3-2-1-1}.
	
	\item
	\textbf{Downlink communication signal model:}
	The signal $\mathbf{s}(t)$ simultaneously propagates to $P$ single-antenna communication users.
	Similar to the received signal models \eqref{eq:3-1-3-2} and \eqref{eq:3-2-3-1}, the frequency-domain signal at the $m$-th subcarrier of the $n$-th OFDM symbol received by the $p$-th user is then expressed as
	\begin{align}
		y_{p,m,n}^{\mathrm{com}}
		= \sum_{l=0}^{L_p-1} \sum_{m'=0}^{M-1}
		\tilde{h}_{p,l}^{\mathrm{com}}\, x_{m',n}\,
		e^{j2\pi \nu_{p,l}^{\mathrm{com}} nT_{\mathrm{sym}}}\,
		e^{-j2\pi m'\Delta f \tau_{p,l}^{\mathrm{com}}}
		\cdot D_M\!\left(m'-m+\epsilon_{p,l}^{\mathrm{com}}\right)
		+ z_{p,m,n}^{\mathrm{com}},
		\label{eq:ofdm-com}
	\end{align}
	where $\tilde{h}_{p,l}^{\mathrm{com}} = h_{p,l}^{\mathrm{com}}\,\mathbf{a}_t^H(\theta_{p,l}^{\mathrm{com}},\phi_{p,l}^{\mathrm{com}})\mathbf{w}e^{j2\pi\nu_{p,l}^{\mathrm{com}} T_{\mathrm{cp}}}$ denotes the effective path gain of the $l$-th path for the $p$-th user, and $\tau_{p,l}^{\mathrm{com}}$ and $\nu_{p,l}^{\mathrm{com}}$ denote the corresponding propagation delay and Doppler frequency; $\epsilon_{p,l}^{\mathrm{com}}=\nu_{p,l}^{\mathrm{com}}/\Delta f$ is the normalized Doppler shift; $L_p$ and $z_{p,m,n}^{\mathrm{com}}$ denote the number of paths and AWGN for the $p$-th user.
	
	\item
	\textbf{Transmitted waveforms of BDs:}
	Similar to Section 4.3.2, PSK is selected as a representative example to elaborate on the transmitted waveforms of BDs.
	The $q$-th BD embeds the identity information by modulating its reflection coefficient in synchronization with the incident OFDM symbols, where one BD symbol is mapped to one OFDM symbol period.
	The transmitted waveform of the $q$-th BD is written as
	\begin{align}
		b_q(t) = \sum_{n=0}^{N-1} \omega_{q,n}\,
		\operatorname{rect}\!\left(\frac{t - nT_{\mathrm{sym}}}{T_{\mathrm{sym}}}\right),
		\label{eq:ofdm-bd}
	\end{align}
	where $\omega_{q,n}$ denotes the $n$-th transmitted BD symbol of the $q$-th BD.
	
	\item
	\textbf{Echo signal model:}
	The echo from each aircraft-BD consists of two components: the structural reflection arising from the physical surface of the aircraft-BD, and the antenna-mode backscatter component carrying the identity information.
	Then, the received echo signal at the ISCAI station is expressed as
	\begin{align}
		\mathbf{r}^{\mathrm{sen}}(t)= \sum_{q=0}^{Q-1}
		\mathbf{a}_r(\theta_q,\phi_q) h_q^{\mathrm{sen}}( \beta_q^{\mathrm{bd}} b_q( t-\tau_q) +\beta_q^{\mathrm{str}})\,e^{j2\pi \nu_{q} t}\,
		\mathbf{a}_t^H(\theta_{q},\phi_{q})\mathbf{s}(t-\tau_q) + \mathbf{z}^{\mathrm{sen}}(t),
		\label{eq:ofdm-echo-cont}
	\end{align}
	where the definitions of $\{h_q^{\mathrm{sen}},\theta_q,\phi_q,\tau_q,\nu_{q},\beta_q^{\mathrm{bd}},\beta_q^{\mathrm{str}}\}$ are the same as those in \eqref{eq:fmcw-echo}, and $\mathbf{z}^{\mathrm{sen}}(t)$ is the equivalent noise including the AWGN and the residual interference following self-interference and clutter suppression.
	Then, $\mathbf{r}^{\mathrm{sen}}(t)$ is sampled at an interval of $1/(M\Delta f)$.
	After applying CP removal and an $M$-point discrete Fourier transform along the fast-time index, the frequency-domain signal on the $m$-th subcarrier of the $n$-th symbol $\mathbf{y}_{m,n}^{\mathrm{sen}}$ is expressed as
	\begin{align}
		\mathbf{y}_{m,n}^{\mathrm{sen}}
		={}& \sum_{q=0}^{Q-1} \sum_{m'=0}^{M-1}
		h_q^{\mathrm{sen}} e^{j2\pi\nu_qT_{\mathrm{cp}}}\,\mathbf{a}_r(\theta_q,\phi_q)\,
		\bigl(\mathbf{a}_t^H(\theta_q,\phi_q)\mathbf{w}\bigr)
		\notag \\
		&\quad\times\bigl(\beta_q^{\mathrm{bd}} \omega_{q,n} + \beta_q^{\mathrm{str}}\bigr)\,
		x_{m',n}\,
		e^{j2\pi \nu_q nT_{\mathrm{sym}}}\,
		e^{-j2\pi m'\Delta f \tau_q}
		D_M\!\left(m'-m+\epsilon_q\right)
		+ \mathbf{z}_{m,n}^{\mathrm{sen}},
		\label{eq:ofdm-td}
	\end{align}
	where $\epsilon_{q}=\nu_{q}/\Delta f$ is the normalized Doppler shift, and $\mathbf{z}_{m,n}^{\mathrm{sen}}$ represents the equivalent noise sample vector.
\end{itemize}

Comparing~\eqref{eq:ofdm-td} with the conventional OFDM-enabled ISAC echo model in~\eqref{eq:3-2-2-1}, the symbol-dependent factor $(\beta_q^{\mathrm{bd}}\omega_{q,n}+\beta_q^{\mathrm{str}})$ is the difference introduced by the identification function.
When the BD stops transmitting its identity sequence and holds its reflection state fixed, this factor becomes a constant complex reflection coefficient that can be absorbed into an effective target gain, and~\eqref{eq:ofdm-td} reduces to~\eqref{eq:3-2-2-1}.

\subsubsection{Multifunctional receiver design}
\begin{itemize}
	\item
	\textbf{Communication receiver design:} 
	The expression~\eqref{eq:ofdm-com} characterizes the received signal model when the communication user is located far from the aircraft-BDs. 
	In this case, since the BDs lack power amplification capability, the backscattered signals from the aircraft-BDs arrive at the communication user with power far below that of the direct link from the ISCAI station, and can therefore be treated as part of the noise. 
	Under this condition, the $p$-th user can directly employ the OFDM channel estimation and data demodulation algorithms reviewed in Section~3.2.3 to recover the transmitted data. 
	When the communication user is in close proximity to an aircraft-BD, however, the reflected signal from the BD becomes non-negligible. 
	Since this reflected signal carries BD-modulated identity symbols, its mathematical structure differs from that of standard multipath components: whereas standard multipath components involve only delay, Doppler shift, and path gain, the BD-reflected signal introduces an additional discrete modulation symbol term. 
	This structural discrepancy prevents conventional channel estimation algorithms from correctly modeling the BD-reflected component, thereby causing interference to data demodulation at the communication user. 
	How to effectively handle such interference at the communication receiver constitutes a problem worthy of further investigation in ISCAI systems.
	For example,~\cite{re:isabc-intfree-2024} adopts FSK-based backscatter modulation and lets the base station actively empty the corresponding subcarriers, so that the frequency-shifted backscattered signal and the primary signal always remain orthogonal in the frequency domain.
	\cite{re:isabc-afdm-2026} instead employs OFDM and AFDM for the primary and the backscattered signals, respectively, thereby avoiding their mutual interference through waveform-domain orthogonality.
	
	\item
	\textbf{Joint sensing and identification receiver design:} 
	Based on \eqref{eq:ofdm-td}, the ISCAI station also needs to design a joint sensing and identification signal processing framework that operates on $\{\mathbf{y}_{m,n}^{\mathrm{sen}}\}$ to simultaneously estimate parameters $\{\theta_q,\phi_q,\tau_q,\nu_q\}$, while detecting the BD symbols $\{\omega_{q,n}\}$ for identity acquisition~\cite{re:lae-iscai-2023-2,re:lae-iscai-2024-3,re:lae-iscai-2026-6}.
	Compared with the received signal model of OFDM-enabled ISAC in~\eqref{eq:3-2-2-1} of Section~3.2.2, the integration of identification introduces additional BD symbol terms $\{(\beta_q^{\mathrm{bd}}\omega_{q,n}+\beta_q^{\mathrm{str}})\}$ into $\mathbf{y}_{m,n}^{\mathrm{sen}}$, which poses the following core challenges. 
	As can be seen from~\eqref{eq:ofdm-td}, the BD symbol $\omega_{q,n}$ is tightly coupled in the frequency domain with the aircraft's delay and Doppler shift parameters as well as with the transmitted data $\{x_{m,n}\}$, exhibiting a nonlinear intertwined structure among the three. Although the transmitted data is fully known at the ISCAI station, the ICI induced by the high-speed movement of aircraft prevents the known data from being directly cancelled via simple per-subcarrier division, thereby increasing the aforementioned coupling complexity. 
	Moreover, the derivation of~\eqref{eq:ofdm-td} is based on the assumption that BDs are perfectly synchronized with the incident OFDM signals, whereas in practice, the low-power design of the BDs limits synchronization precision, and the internal clock generator suffers from frequency offsets and timing drifts. 
	As a result, a single BD symbol may straddle two adjacent OFDM symbols, introducing ICI and invalidating the ideal signal model~\eqref{eq:ofdm-td}, which imposes additional challenges on joint sensing and identification receiver design.
\end{itemize}

Several approaches to the joint recovery of parameters and symbols have been reported.
\cite{re:isabc-qintao-letter-2024} decouples the two-dimensional off-grid SBL into two one-dimensional problems, thereby avoiding the prohibitive dictionary size incurred by a joint two-dimensional grid when multiple devices coexist.
\cite{re:lae-iscai-2024-3} instead separates the two tasks into two stages, where the AoAs and delays of the BDs are first estimated by an off-grid SBL-based algorithm, and the channel gain of the backscattered link is then estimated by the LS method to detect the symbols transmitted by the BDs.
Following a similar dictionary-based formulation,~\cite{re:isabc-shao-2024} mounts the reflective modulator directly on the device to be localized, and casts localization and information recovery as a compressive sensing problem that is solved jointly by a message passing algorithm.
\cite{re:isabc-uavswarm-2026} lets a leading UAV act as the primary transmitter while the remaining swarm UAVs act as passive BDs, and adopts 3D sparse nested arrays at the base station to obtain the angular resolution required for sensing the densely located swarm UAVs.
The above works consider static scenarios, where the parameters to be recovered are the angles and, in some cases, the delays.
In addition, the structural reflection component is absent from their echo models.

\cite{re:lae-iscai-2023-2} and~\cite{re:lae-iscai-2026-1} further investigate the joint recovery problem in moving scenarios.
To this end, in~\cite{re:lae-iscai-2023-2}, a transmission frame structure of the BDs is designed to solve the coupling issue, where each frame starts with a pilot symbol so that the target parameters can be estimated from the pilot positions independently of the random BD symbols, and the 3D-Newtonized OMP estimator is then utilized to realize joint parameter estimation and BD symbol detection.
In contrast, the work of~\cite{re:lae-iscai-2026-1} proposes a novel iterative detection and estimation (IDE) framework which involves the following two processes alternately.
Parameter estimation is first performed using an improved ML estimator based on detected BD symbols, and then BD symbols are detected using an ML detector based on estimated target parameters. 
The two works treat the ICI differently: it is regarded as noise in~\cite{re:lae-iscai-2023-2}, which limits the achievable delay and Doppler accuracy, whereas it is exploited in~\cite{re:lae-iscai-2026-1} to construct the initial estimates.

\subsubsection{Mutual integration gains}

In ISCAI, communication, sensing, and identification achieve comprehensive spectrum and hardware sharing while enabling automatic association of targets' physical parameters and identity information.
Their deep integration brings the following significant mutual integration gains. 
\begin{itemize}
	\item
	ISCAI retains the four mutual integration gains of ISAI in Section~\ref{sec:4-3-3}. 
	First, BDs can reuse communication and sensing signals as RF carriers for BackCom to complete identification. 
	Second, the introduction of BDs enlarges the equivalent RCS of aircraft to enhance sensing accuracy. 
	As shown in~\cite{re:lae-iscai-2026-1}, the Miller--Chang bounds for delay and Doppler shift estimation are effectively decreased after the introduction of BDs, thereby improving the achievable sensing accuracy. 
	Third, identification results can directly guide sensing resource allocation. Fourth, identity information can facilitate data association in distributed architectures.
	
	\item
	The high-precision 3D locations acquired by sensing and the identity information acquired by identification can jointly guide communication resource scheduling.
	For example, the system can prioritize allocating communication resources to aircraft identified as authorized and improve transmission rates and anti-interference capability through precise beam steering based on sensed locations.
	Furthermore, predicted flight trajectories derived from continuous sensing updates enable proactive handover management and spectrum pre-allocation, reducing handover latency and interruption.
	
	\item
	In addition, uplink communication signals can also carry identity information and onboard sensor data to further assist identification and environmental sensing.
	Specifically, aircraft can report onboard measurements such as barometric altitude, coordinates, and airspeed via uplink, which serve as cross-validation references to improve identification reliability.
	The aggregated sensor data from multiple aircraft enables cooperative environmental sensing, such as constructing distributed radio maps, which in turn enhances the situational awareness and sensing coverage.
\end{itemize}

\begin{table*}[!t]
	\centering
	\footnotesize
	\setlength{\tabcolsep}{4pt}
	\renewcommand{\arraystretch}{1.15}
	\caption{Systematic comparison among ISAC, ISAI, and ISCAI}
	\label{tab:isac-isai-iscai-comparison}
	\begin{tabularx}{\textwidth}{
			@{}
			>{\raggedright\arraybackslash}p{0.13\textwidth}
			>{\raggedright\arraybackslash}X
			>{\raggedright\arraybackslash}X
			>{\raggedright\arraybackslash}X
			@{}
		}
		\toprule
		Comparison dimension
		& ISAC
		& ISAI
		& ISCAI\\
		\midrule
		
		Integrated functionalities
		& Communication $+$ sensing
		& Sensing $+$ identification
		& Communication $+$ sensing $+$ identification \\
		
		\addlinespace[1pt]
		Design objective
		& Data transmission and target physical-state estimation: ``where the target is and how it moves.''
		& Target physical-state estimation and individual identity acquisition: ``where the target is and who it is,'' without communication services.
		& Data transmission, physical-state estimation, and identity acquisition, supporting the detection--estimation--identification--control loop. \\
		
		\addlinespace[1pt]
		Multifunctional signal
		& Communication-centric: OFDM, OTFS, and AFDM; sensing-centric: data-embedded FMCW (Sections~3.2 and~3.3).
		& Sensing waveforms such as FMCW also serve as BD RF excitation signals (dual-functional; Section~4.3.2).
		& Communication waveforms such as OFDM also serve as sensing and BD RF excitation signals (triple-functional; Section~4.4.2). \\
		
		\addlinespace[1pt]
		Receiver outputs
		& Sensing: estimate $\{\tau,\nu,\theta,\phi\}$; communication: channel estimation and data demodulation.
		& Jointly estimate $\{\tau,\nu,\theta,\phi\}$ and detect BD identity symbols $\{\omega_{q,k,l}\}$ (Section~4.3.3).
		& Users recover communication data; the station jointly estimates $\{\tau,\nu,\theta,\phi\}$ and detects $\{\omega_{q,n}\}$ (Section~4.4.3). \\
		
		\addlinespace[1pt]
		Representative performance metrics
		& Rate and BER; detection probability, MSE, and CRLB.
		& Sensing metrics; identification accuracy, missed-identification probability, false-identification probability, and latency.
		& Communication, sensing, and identification performance metrics. \\
		
		\addlinespace[1pt]
		Representative application scenarios
		& Integrated communication and physical-state sensing, e.g., aircraft monitoring and control together with trajectory tracking.
		& Surveillance without communication services, e.g., controlled-area monitoring and low-cost upgrades of existing FMCW radar platforms.
		& Dense airspace requiring communication, continuous tracking, and real-time identity acquisition, e.g., urban logistics networks and UAM corridors. \\
		
		\addlinespace[1pt]
		Functional relationship
		& \multicolumn{3}{
			>{\raggedright\arraybackslash}p{0.80\textwidth}
		}{
			ISCAI $\rightarrow$ ISAI when communication is disabled; ISCAI $\rightarrow$ ISAC when identification is disabled; and ISAI $\rightarrow$ conventional sensing when identification is disabled.
		} \\
		
		\bottomrule
	\end{tabularx}
	
	\vspace{2pt}
	\begin{minipage}{\textwidth}
		\raggedright
		\footnotesize
		\textit{Note:} For the ISAI and ISCAI columns, the multifunctional signals and receiver outputs are presented according to the BackCom-based implementations adopted in this paper.
	\end{minipage}
\end{table*}

\subsection{Summary}
This section introduced the ISCAI paradigm for LAINs. 
We first discussed the evolution from ISAC to ISCAI, and then elaborated the BackCom-based identification paradigm, including the basic principles of BackCom, identification process, identification performance metrics, as well as main advantages and limitations. 
Building on this, we further introduced the ISAI and ISCAI paradigms, and explored their system models, signal models, multifunctional receiver design, and mutual integration gains. 
To further clarify the distinctions and relationships among the three integrated frameworks, Table~\ref{tab:isac-isai-iscai-comparison} systematically compares ISAC, ISAI, and ISCAI.

\section{Open problems and future directions}
\label{sec:ISCAI-open}
Although ISCAI offers more advantages than ISAC for LAINs, the integration of identification also brings several intrinsic challenges, such as performance tradeoffs and theoretical bounds among communication capacity, sensing accuracy, and identification accuracy, as well as complex system design.
To fully unleash the potential of ISCAI for LAINs, we elaborate on three open problems and future research directions in this field, including the information-theoretic framework of ISCAI, AI-empowered ISCAI system design, and RIS-enhanced ISCAI.

\subsection{Information-theoretic framework of ISCAI}

Establishing a unified information-theoretic framework for ISCAI to characterize the performance interdependencies and joint Pareto boundaries among communication, sensing, and identification is the theoretical foundation for guiding system designs such as waveform optimization and resource allocation.
However, the three functions have inherent mismatches in performance metrics~\cite{re:lae-future-information-2022-1,re:on-the-funda-fanliu-2023}: communication is typically measured by Shannon channel capacity or data rate, sensing is measured by the detection probability or CRLB, and identification is measured by the reliability and timeliness metrics defined in Section 4.2.3.
This heterogeneous metric system significantly increases the difficulty of constructing a unified information-theoretic framework.

For conventional ISAC, the communication--sensing tradeoff has been characterized by the CRLB--rate region in~\cite{re:on-the-funda-fanliu-2023}, whose boundary can be obtained by maximizing the communication rate subject to a prescribed sensing-distortion constraint.
Let
$\boldsymbol{\eta}
=
[\boldsymbol{\eta}_0^T,\ldots,
\boldsymbol{\eta}_{Q-1}^T]^T$,
where
$\boldsymbol{\eta}_q
=
[\tau_q,\nu_q,\theta_q,\phi_q]^T$
collects the continuous sensing parameters of the $q$-th aircraft-BD. Moreover, let $\mathbf{R}_s$ denote the covariance matrix of the transmitted ISAC signal $\mathbf{s}(t)$, and let $\mathcal{P}_{\max}$ denote the transmit power budget. 
A representative covariance-domain formulation is
\begin{align}
	\max_{\substack{
			\mathbf{R}_s\succeq\mathbf{0}:\\
			\operatorname{tr}(\mathbf{R}_s)
			\leq\mathcal{P}_{\max}}}
	R_{\mathrm{com}}(\mathbf{R}_s)
	\quad
	\mathrm{s.t.}\quad
	\operatorname{tr}
	\left[
	\mathbf{J}_{\boldsymbol{\eta}}^{-1}
	(\mathbf{R}_s)
	\right]
	\leq
	\mathcal{D}_{\mathrm{sen}},
	\label{eq:isac-cd}
\end{align}
where $R_{\mathrm{com}}(\mathbf{R}_s)$ denotes the achievable communication rate, $\mathbf{J}_{\boldsymbol{\eta}}(\mathbf{R}_s)$ is the FIM of $\boldsymbol{\eta}$, and $\mathcal{D}_{\mathrm{sen}}$ is the allowable sensing distortion. 
Varying $\mathcal{D}_{\mathrm{sen}}$ traces a CRLB--rate boundary of ISAC.

Extending~\eqref{eq:isac-cd} to ISCAI first requires an appropriate identification-distortion metric. Since aircraft identity is a discrete variable, a conventional CRLB cannot be directly applied to it. Let $W_{\mathrm{id}}$ and $\hat W_{\mathrm{id}}$ denote the true and estimated aircraft identities, respectively. 
By adopting the Hamming distortion $d_{\mathrm{id}}	(W_{\mathrm{id}},\hat W_{\mathrm{id}})	=	\mathbf{1}	\{W_{\mathrm{id}}\neq\hat W_{\mathrm{id}}\}$, the average identification distortion is
\begin{align}
	\mathcal{D}_{\mathrm{id}}
	=
	\mathbb{E}
	\left[
	d_{\mathrm{id}}
		(W_{\mathrm{id}},\hat W_{\mathrm{id}})
		\right]
		=
		1-A_{\mathrm{id}},
		\label{eq:id-distortion}
	\end{align}
where $A_{\mathrm{id}}$ denotes the identification accuracy.
Based on the above definition, the rate--sensing--identification tradeoff among communication rate, sensing accuracy, and identification reliability can be characterized by the following constrained optimization problem:
\begin{align}
		\max_{\substack{
				\mathbf{R}_s\succeq\mathbf{0},\,
				\mathcal{M}_{\mathrm{id}}:\\
				\operatorname{tr}(\mathbf{R}_s)
				\leq\mathcal{P}_{\max}}}
		& R_{\mathrm{com}}
		(\mathbf{R}_s,\mathcal{M}_{\mathrm{id}})
		\qquad
		\mathrm{s.t.}\quad
		\operatorname{tr}
		\left[
		\left(
		\mathbf{J}_{\boldsymbol{\eta}}^{\mathrm{eff}}
		\right)^{-1}
		\right]
		\leq
		\mathcal{D}_{\mathrm{sen}},
		\quad
		1-A_{\mathrm{id}}
		(\mathbf{R}_s,\mathcal{M}_{\mathrm{id}})
		\leq
		\mathcal{D}_{\mathrm{id}},
		\label{eq:iscai-cd}
	\end{align}
where $\mathcal{M}_{\mathrm{id}}$ collectively represents the BD identity-codebook and symbol-mapping design, and $\mathbf{J}_{\boldsymbol{\eta}}^{\mathrm{eff}}$ denotes the effective FIM of the continuous sensing parameters in the presence of identity uncertainty. 
When identity transmission is disabled and the BD reflection state is fixed,~\eqref{eq:iscai-cd} reduces to the conventional ISAC formulation in~\eqref{eq:isac-cd}. 
Fixing $\mathcal{D}_{\mathrm{id}}$ yields an identification-constrained CRLB--rate curve, whereas jointly varying $\mathcal{D}_{\mathrm{sen}}$ and $\mathcal{D}_{\mathrm{id}}$ traces a 3D Pareto surface.
An obvious theoretical difficulty is that the sensing parameters are continuous, whereas the aircraft identities and BD symbols are discrete. 
Therefore, $\mathbf{J}_{\boldsymbol{\eta}}^{\mathrm{eff}}$ cannot be obtained by directly treating the identity as an ordinary continuous nuisance parameter.

\subsection{AI-empowered ISCAI system design}

Faced with complex and dynamic low-altitude environments and the drastically increased variable dimensionality of ISCAI systems, conventional system design methods based on precise analytical models suffer from intractable solutions, prohibitive computational complexity, and poor environmental adaptability. 
Introducing AI techniques such as deep learning and reinforcement learning, through data-driven or hybrid data- and model-driven paradigms~\cite{re:lae-future-ai-2024-1,re:lae-spe-2026-8,re:lae-future-ai-2026-2,re:lae-future-ai-2026-3}, provides new solution paths for the highly nonlinear joint design problems in ISCAI systems.

In terms of waveform design, unlike conventional ISAC where the tradeoff involves only communication and sensing, ISCAI waveforms must simultaneously balance the performance requirements of communication rate, sensing accuracy, and identification accuracy, for which closed-form optimal solutions are generally unavailable. 
A further difficulty is specific to ISCAI: since BDs are passive, identification is realized only when the incident power at a BD exceeds the activation threshold of its energy harvesting circuit. 
This constraint is non-differentiable with respect to the transmit covariance and becomes combinatorial when multiple aircraft-BDs are present. 
Deep reinforcement learning can be leveraged to construct an adaptive waveform design framework, in which the state consists of the instantaneous channel states and the estimated aircraft distribution, the action consists of the transmit covariance and the reflection constellation parameters of the BDs, and the reward is composed of the three metrics introduced in Section 5.1, i.e.,
$r=w_1R_{\mathrm{com}}-w_2\,\operatorname{tr}\big\{\big(\mathbf{J}_{\boldsymbol{\eta}}^{\mathrm{eff}}\big)^{-1}\big\}-w_3\mathcal{D}_{\mathrm{id}}$, where $w_1$, $w_2$, and $w_3$ are weighting factors.

In terms of receiver design, communication symbols, aircraft parameters, and BD symbol sequences exhibit deep nonlinear coupling in the received signal, as reflected by the bilinear structure of~\eqref{eq:fmcw-beat-discrete} and~\eqref{eq:ofdm-td} discussed in Sections~4.3.3 and~4.4.3, which makes it difficult for conventional analytical methods to balance computational complexity and multifunctional extraction accuracy.
To this end, advanced AI techniques such as model-driven deep unfolding can be employed to directly learn the nonlinear inverse mapping from complex echo signals to communication symbols, sensing parameters, and identity information.
An example is to unfold the IDE framework reviewed in Section 4.4.3 into a network with a fixed number of layers, in which each layer corresponds to one alternation between parameter estimation and symbol detection, and the damping factors, soft-decision thresholds, and initialization are learned from data.

\subsection{RIS-enhanced ISCAI}

RIS is a planar metamaterial array composed of a large number of low-cost controllable electromagnetic reflection elements, and is capable of achieving the intelligent reconfiguration of wireless channels by adjusting the phases and amplitudes of incident electromagnetic waves through controllers~\cite{re:lae-ris-2020-2, re:lae-ris-2021-1,re:lae-ris-2025-7}.
In ISCAI, RIS affects not only the station--user communication link but also the station--aircraft-BD--station sensing and identification link. 
First, RIS can be deployed on building surfaces as a signal enhancer, using beamforming to enhance communication links while strengthening target echoes to improve sensing and identification performance.
Second, RIS can serve as a clutter suppressor by generating destructive interference in specific directions through reflection coefficient control, suppressing strong clutter from static scatterers such as buildings.
Third, RIS can expand the coverage ranges of ISCAI systems by constructing virtual LoS links, thereby achieving communication for users or sensing and identification for targets within occluded areas. 
Let $\boldsymbol{\Phi}$ denote the diagonal RIS reflection matrix and let $\mathbf{w}$ denote the transmit beamforming vector. 
A representative joint beamforming and RIS configuration problem can be formulated as
\begin{align}
	\max_{\mathbf{w},\boldsymbol{\Phi}}
	\quad
	&
	R_{\mathrm{com}}
	(\mathbf{w},\boldsymbol{\Phi})
	\nonumber\\
	\mathrm{s.t.}\quad
	&
	\operatorname{tr}
	\left[
	\left(
	\mathbf{J}_{\boldsymbol{\eta}}^{\mathrm{eff}}
	(\mathbf{w},\boldsymbol{\Phi})
	\right)^{-1}
	\right]
	\leq
	\mathcal{D}_{\mathrm{sen}},\quad
	1-A_{\mathrm{id}}
	(\mathbf{w},\boldsymbol{\Phi})
	\leq
	\mathcal{D}_{\mathrm{id}},\quad
	\mathcal{P}_{\mathrm{BD}}^{\mathrm{in}}
	(\mathbf{w},\boldsymbol{\Phi})
	\geq
	\mathcal{P}_{\mathrm{BD}}^{\mathrm{act}},
	\quad
	\left|
	[\boldsymbol{\Phi}]_{r,r}
	\right|
	=
	1,
	\label{eq:ris-iscai-optimization}
\end{align}
where $\mathcal{P}_{\mathrm{BD}}^{\mathrm{in}}$ is the incident power received by the BD; $\mathcal{P}_{\mathrm{BD}}^{\mathrm{act}}$ is the activation threshold of its energy-harvesting circuit; and $[\boldsymbol{\Phi}]_{r,r}$ denotes the reflection coefficient of the $r$-th RIS element.
The activation constraint is important for ISCAI because identification becomes infeasible when the passive BD receives insufficient incident power, rather than merely suffering a gradual performance degradation.
Solving~\eqref{eq:ris-iscai-optimization} requires the joint acquisition of the station--RIS, RIS--user, and RIS--aircraft-BD channel states. 
In highly mobile low-altitude environments, these cascaded channels may become outdated before the RIS configuration is applied. 
Future research should investigate mobility-aware channel prediction, low-overhead RIS training, and robust joint beamforming and phase optimization under imperfect channel and aircraft-state information.

\section{Conclusions}
\label{sec:con}
The rapid development of LAE is driving LAINs to evolve beyond connectivity-oriented designs toward a more integrated support of communication, sensing, and identification. 
This paper has presented a comprehensive review of ISCAI for LAE. 
A central observation of this survey is that the evolution from ISAC to ISCAI requires more than placing an existing identity-reporting system alongside ISAC. 
The identification mechanism must be compatible with the ISAC waveform and echo-processing chain so that target states and individual identities can be jointly acquired and directly associated.
We identify BackCom-based electronic license plates as a key enabling route toward this objective and use this route to organize the resulting ISAI and ISCAI research framework.
We first examined the main standalone approaches to low-altitude communication, sensing, and identification, and clarified how representative application scenarios lead to their integration into a unified design framework. 
We then reviewed the fundamentals of low-altitude ISAC, including system architectures, channel modeling, unified ISAC signal models, and performance metrics, followed by communication-centric and sensing-centric ISAC system design, as well as resource allocation and aircraft trajectory optimization. 
Building on this foundation, we discussed the evolution from ISAC to ISCAI, and elaborated on the BackCom-based identification paradigm in which BDs serve as electronic license plates for aircraft. We further presented the system and signal models of the resulting ISAI and ISCAI paradigms, along with their multifunctional receiver design and mutual integration gains. 
Finally, several open challenges and future research directions have been discussed, including the information-theoretic framework of ISCAI, AI-empowered ISCAI system design, and RIS-enhanced ISCAI. 
These insights aim to guide future research and practical deployments toward more capable, intelligent, and safe LAINs.

\Acknowledgements{This work was supported in part by the National Natural Science Foundation of China under Grants U24A20210 and 62501120, in part by the Program of Introducing Talents of Discipline to Universities under Grant B20064, in part by the Postdoctoral Fellowship Program of China Postdoctoral Science Foundation under Grant GZC20252303, and in part by the China Postdoctoral Science Foundation under Grant 2025M773495.}

\bibliographystyle{scis}
\bibliography{scis-reference}

\end{document}